\documentclass[a4paper,11pt]{article}
\pdfoutput=1 % to allow compilation in JCAP
\usepackage[T1]{fontenc} % use actual unicode characters in PDF
\usepackage{lmodern} % scaleable font (to run with fontenc)
\usepackage{jcappub}
\usepackage{amsmath}
\usepackage{amssymb}
\usepackage{bm}
\usepackage{graphicx}
\usepackage{enumitem}
\usepackage{geometry}
\usepackage{pdflscape}
\usepackage{enumerate}
\usepackage{xspace} %Allows definitions to have spaces after them more easily
\usepackage{soul}
\usepackage{accents}

\usepackage{mathtools}

\makeatletter
\gdef\@fpheader{}
\g@addto@macro\bfseries{\boldmath}
\makeatother

\newcommand{\ie}{\textsl{i.e.~}}

\newcommand{\eg}{\textsl{e.g.~}}

\newcommand{\dd}{\mathrm{d}}
\newcommand{\Rea}{\Re \mathrm{e}\,}

\newcommand{\efolds}{$e$-folds}

\newcommand{\beq}{\begin{equation}}
\newcommand{\eeq}{\end{equation}}
\newcommand{\bea}{\begin{equation}\begin{aligned}}
\newcommand{\eea}{\end{aligned}\end{equation}}

\newlength{\wsingfig}
\newlength{\wdblefig}
\newlength{\wquadfig}
\newlength{\wtriplefig}
\newcommand{\Eqs}[1]{Eqs.~(\ref{#1})}

\newcommand{\Refa}[1]{Ref.~{\cite{#1}}}

\newcommand{\Sec}[1]{Sec.~\ref{#1}}

\newcommand{\App}[1]{Appendix~\ref{#1}}

\newcommand{\eq}[1]{Eq.~(\ref{#1})}
\newcommand{\fig}[1]{Fig.~\ref{#1}}
\newcommand{\figs}[1]{Figs.~\ref{#1}}

\newcommand{\Nb}{{\mathcal N}}
\newcommand{\piphi}{{\pi_\phi}}

\def\N{\mathcal{N}}
\def\R{\mathcal{R}}

\def\init{\mathrm{in}}

\def\Nmean{\langle \mathcal{N} \rangle}

\def\d{{\mathrm d}}

\def\figurewidth{0.85}
\def\halffigurewidth{0.48}

\def\pyfpt{\textsc{PyFPT}\xspace}

\def\efolds{e-folds\xspace}
\def\efolding{e-folding\xspace}

\usepackage{xcolor}

\subheader{}

\title{
Stochastic inflation beyond slow roll: noise modelling and importance sampling
}

\hypersetup{pdftitle={Stochastic inflation beyond slow roll: noise modelling and importance sampling},pdfauthor={6 authors}}

\author[a]{Joseph H.~P.~Jackson,}
\author[a,b]{Hooshyar Assadullahi,}
\author[a]{Andrew D.~Gow,}
\author[a]{Kazuya Koyama,}
\author[c,a]{Vincent Vennin,}
\author[a]{David Wands}

\affiliation[a]{Institute of Cosmology \& Gravitation, University of Portsmouth, Dennis Sciama Building, Burnaby Road, Portsmouth, PO1 3FX, United Kingdom}
\affiliation[b]{School of Mathematics and Physics, University of Portsmouth, Lion Gate Building, Lion Terrace, Portsmouth, PO1 3HF, United Kingdom}
\affiliation[c]{Laboratoire de Physique de l'École Normale Supérieure, ENS, CNRS, Université PSL, Sorbonne Université, Université Paris Cité, F-75005 Paris, France}

\emailAdd{joseph.jackson@port.ac.uk}
\emailAdd{hooshyar.assadullahi@port.ac.uk}
\emailAdd{andrew.gow@port.ac.uk}
\emailAdd{kazuya.koyama@port.ac.uk}
\emailAdd{vincent.vennin@ens.fr}
\emailAdd{david.wands@port.ac.uk}

\date{today}

\begin{document}

\sloppy

\abstract{We simulate the distribution of very rare, large excursions in the primordial density field produced in models of inflation in the very early universe which include a strong enhancement of the power spectrum. The stochastic $\delta \Nb$ formalism is used to identify the probability distribution for the primordial curvature perturbation with the first-passage-time distribution, $P(\delta \Nb)$, and we compare our stochastic results with those obtained in the classical $\delta \Nb$ approach. We extend the \pyfpt numerical code to simulate the full 2D phase space, and apply importance sampling which allows very rare fluctuations to be simulated in $\mathcal{O}$(10) minutes on a single CPU, where previous direct simulations required supercomputers. We demonstrate that the stochastic noise due to quantum fluctuations after a sudden transition to ultra-slow roll can be accurately modelled using an analytical Bessel-function ansatz to identify the homogeneous growing mode. The stochastic noise found in this way is a function of the field value only. This enables us to coarse grain the inflation field at the Hubble scale and include non-linear, stochastic evolution on all super-Hubble length scales.
}

\keywords{Cosmological perturbation theory, inflation, physics of the early universe, primordial black holes}

% \arxivnumber{XXXX.XXXXX}

\maketitle

\section{Introduction}

Inflation is a period of accelerated expansion in the very early universe which can explain the observed isotropy, homogeneity and flatness of our Universe~\cite{Starobinsky:1980te, Sato:1980yn, Guth:1980zm, Linde:1981mu, Albrecht:1982wi, Linde:1983gd}. Its prediction of an almost scale-invariant spectrum of primordial perturbations is in excellent agreement with current Cosmic Microwave Background (CMB) observations~\cite{Planck2018,Tristram:2021tvh}. These observations constrain the distribution of density fluctuations, which originate from quantum vacuum fluctuations during inflation but are stretched to macroscopic scales, thereby constraining the dynamics of inflation on large scales~\cite{Mukhanov:1981xt, Mukhanov:1982nu, Starobinsky:1982ee, Guth:1982ec, Hawking:1982cz, Bardeen:1983qw}. 

There are fewer constraints on smaller scales during inflation, leading to much interest in the possibility of larger perturbations on these scales~\cite{Chluba:2015bqa}. Sufficiently large  over-densities can collapse post-inflation to form primordial black holes (PBHs)~~\cite{Zeldovich1967, Hawking:1971ei, Carr:1974nx}. These are a dark matter candidate~\cite{Carr:2016drx, Carr:2020xqk, Green:2020jor, Green:2024bam} and could possibly explain several other observational puzzles as well~\cite{Carr:2019kxo}. Large scalar perturbations can also source a stochastic background of gravitational waves at second order in perturbation theory~\cite{Ananda:2006af, Baumann:2007zm} and therefore possibly explain the recent pulsar-timing array detections~\cite{EPTA:2023xxk,NANOGrav:2023hvm}.

Very large and rare perturbations are required to produce a measurable abundance of PBHs~\cite{LISACosmologyWorkingGroup:2023njw}. However, such an enhanced power spectrum also brings into question the validity of linear calculations, and so a non-linear treatment of fluctuations is required to reliably calculate PBH abundances~\cite{Pattison:2017mbe}.

Stochastic inflation is an effective-theory approach to modelling non-linear inflationary dynamics~\cite{Starobinsky:1986fx,Starobinsky:1994bd}. This is done by mapping the quantum field theory in curved spacetime onto a stochastic description, splitting the full quantum field into short- and long-wavelength regimes. While perturbative quantum field theory is used to describe the evolution of short-scale modes in a local patch, the long-wavelength regime is modelled using a separate-universe approach, where above the coarse-graining scale the fields obey the evolution equations for a locally homogeneous and isotropic (Friedmann--Lemaître--Robertson--Walker, FLRW) universe. As small-scale quantum fluctuations are stretched and cross the coarse-graining scale, they give a stochastic kick to the long-wavelength fields.

In the $\delta \Nb$ formalism~\cite{Starobinsky:1982ee, Starobinsky:1986fxa, Sasaki1996, Sasaki:1998ug, Lyth:2004gb}, the non-linear primordial density perturbation is identified with the fluctuation in the local duration of inflation in different parts of our Universe. 
In general there is an exponential enhancement in the probability of large perturbations found using the stochastic $\delta \Nb$ formalism compared to linear theory~\cite{Vennin:2015hra, Pattison:2017mbe, Ezquiaga:2019ftu, Animali:2022otk, Briaud:2023eae}.  
However, computing the amplitude of this enhancement can be difficult. Adopting a brute force approach, billions of stochastic simulations are required to find the few rare events with perturbations large enough to produce PBHs, requiring supercomputers to directly sample the expected distribution~\cite{Figueroa:2020jkf, Figueroa:2021zah}.
Here for the first time we apply importance sampling~\cite{Mazonka:1998ge, Jackson:2022unc} to inflationary models which experience a strong enhancement of density perturbations on small scales. With this approach, the events of interest can be directly simulated and their probability found, reducing the number of simulations required by orders of magnitude. What would otherwise require many hours on a supercomputer takes only minutes on a laptop.

We compare the results from our stochastic simulations with those obtained in the classical $\delta \Nb$ approach, where a scalar field perturbation is applied at one single time during inflation and the subsequent non-linear, but classical evolution is followed to the end of inflation. In this case the scalar field fluctuations are computed about a fixed classical background, rather than the stochastically varying local background. Nonetheless, the non-linear evolution subsequently in the long-wavelength limit can lead to an enhancement of the probability of large fluctuations with respect to linear theory~\cite{Biagetti:2018pjj, Firouzjahi:2020jrj, Cai:2022erk, Pi:2022ysn, Hooshangi:2023kss, Wang:2024vfv} and in some cases are expected to give a good description of the stochastic simulations~\cite{Tomberg:2023kli, Ballesteros:2024pwn}. The classical $\delta \Nb$ results can therefore be used to test the stochastic noise model, found using the classical background for a fixed range of modes, and the importance sampling approach proposed here, which is the main focus of this work.

Either formalism uses the separate-universe approach which requires one to identify the long-wavelength limit (what we will refer to as the homogeneous part) of the linearised field perturbation with the inhomogeneous field perturbation at a finite scale, usually at, or soon after, Hubble exit during inflation. Although this is relatively straight-forward during slow roll, it is non-trivial in models with a sudden transition and non-slow-roll evolution~\cite{Jackson:2023obv, Artigas:2024xhc}, such as periods of ultra-slow roll~\cite{Dimopoulos:2017ged}.

In this paper we first introduce the background equations for the separate-universe approach, the $\delta N$ formalism and stochastic inflation in Sec.~\ref{sec:background}. A classical $\delta N$ formula is also derived, for comparison with the stochastic results. In Sec.~\ref{sec:noise_models}, different methods for finding the stochastic noise are discussed. We argue that the noise remains one-dimensional in the two-dimensional phase space, even immediately after the sudden transition. Post-transition we find the noise to be only a function of the field-space position, and not the full trajectory. Numerical results of our stochastic simulations are presented in Sec.~\ref{sec:stochastic_results}, along with details about numerical modelling methods. We present our conclusions and possible directions for future work in Sec.~\ref{sec:conclusion}.

\section{The separate-universe approach and \texorpdfstring{$\delta \Nb$}{δN}}
\label{sec:background}

\subsection{Background phase space and the separate-universe approach}

In canonical single-field inflation, the universe is dominated by a scalar field $\phi$ with a potential $V(\phi)$ at very early times. The Klein--Gordon equation of motion is
\begin{equation}
    \label{eq:klein_gordon}
    \Ddot{\phi} + 3H \dot{\phi} + \frac{\d V(\phi)}{\d \phi} = 0 \, ,
\end{equation}
where a dot denotes a cosmic time derivative. The Hubble rate, $H$, is given by the Friedmann constraint equation
\begin{equation}
\label{eq:Friedmann}
    3 H^2 = V(\phi) + \frac12 \dot\phi^2 \, , 
\end{equation}
where we use natural units, such that $8\pi G=1$.
One way to characterise the dynamics of inflation is to use the Hubble-flow parameters
\begin{equation}
\label{eq:hubble_flow_parameters}
    \epsilon_{i+1} = \frac{\d \ln \epsilon_{i}}{\d N} \quad {\rm where} \quad \epsilon_0 \equiv \frac{H_\init}{H} \, .
\end{equation}
In particular, $\epsilon_1=\dot{\phi}^2/2H^2$, and for inflation to occur one requires $\epsilon_1<1$. 

We will first consider linear perturbation theory to describe scalar field fluctuations during inflation. In the spatially-flat gauge, which we shall use unless otherwise specified, this gives an equation of motion for each Fourier mode with comoving wavenumber $k$~\cite{Bassett:2005xm}
\begin{equation}
\label{eq:full_delta_phi}
        \Ddot{\delta\phi}_{k} + 3H \dot{\delta\phi}_{k} + \left[ \frac{k^2}{a^2} + \frac{\d^2 V}{\d \phi^2} - a^{-3} \frac{\d}{\d t} \left( \frac{a^3\dot\phi^2}{H} \right) \right] \delta\phi_{k} = 0 \, .
\end{equation}
If we quantise the field fluctuations and assume they start in the Bunch--Davies vacuum state at early times during inflation, then we have the particular solution on small scales
\begin{equation}
\label{eq:bunch_davies}
    \delta \phi_{k}^{\mathrm{BD}} = \frac{e^{-ik\eta}}{a\sqrt{2k}} \quad {\rm for} \quad k \gg aH \,,
\end{equation}
where the conformal time $\eta\equiv\int \d t/a$.

During inflation small, sub-Hubble scales ($k\gg aH$) are stretched up to super-Hubble scales ($k\ll aH$). 
In the long-wavelength limit, where $k/aH \rightarrow 0$, \eq{eq:full_delta_phi} reduces to
\begin{equation}
\label{eq:seperate_universe_approach_delta_phi}
        \Ddot{\delta\phi}_{\mathrm{h}} + 3H \dot{\delta\phi}_{\mathrm{h}} + \left[ \frac{\d^2 V}{\d \phi^2} - a^{-3} \frac{\d}{\d t} \left( \frac{a^3\dot\phi^2}{H} \right) \right] \delta\phi_{\mathrm{h}} = 0 \, ,
\end{equation}
which is consistent with the separate-universe approach~\cite{Sasaki:1998ug,Wands:2000dp,Lyth:2004gb,Pattison:2019hef}.
Here, $\delta \phi_{\mathrm{h}}$ is the perturbation to the field value in a locally-homogeneous super-Hubble patch, and it follows the homogeneous behaviour on super-Hubble scales. Specifically, one needs to evolve the local Klein--Gordon equation~\eqref{eq:klein_gordon} using the local Hubble rate obtained from~\eqref{eq:Friedmann}, while allowing for perturbations of the local proper time with respect to the background cosmic time\footnote{The momentum constraint in perturbation theory relates the perturbation in the local lapse function to the local field perturbation, which is needed to relate the local proper time in the separate-universe patch to the global time coordinate being used in \eq{eq:seperate_universe_approach_delta_phi}~\cite{Kodama:1997qw,Sasaki:1998ug,Pattison:2019hef}.}. The locally-homogeneous dynamics are expected to give a good approximation of the full inhomogenous dynamics on sufficiently large scales.

The general solution of \eq{eq:seperate_universe_approach_delta_phi}, in terms of the number of \efolds $N= \ln{\left( a \right)}$, is given by
\begin{equation}
\label{eq:homogeneous_delta_phi}
\delta \phi_{\mathrm{h}} (N) = C \sqrt{2 \epsilon_1 (N)} + D \sqrt{2 \epsilon_1 (N)}  \int^{N_{\mathrm{end}}}_{N} \frac{\d \tilde{N}}{ 2\tilde{a}^3\tilde{H} \tilde{\epsilon}_1} \, ,
\end{equation}
where $C$ and $D$ are constants of integration and $N_\text{end}$ is the number of e-folds at the end of inflation. The first term is known as the \textit{growing} mode and the second as the \textit{decaying} mode. By construction the decaying mode vanishes at the end of inflation.

The comoving curvature perturbation during inflation is given by $\R=H\delta\phi/\dot\phi$ in linear theory\footnote{In this paper $\R$ and $\delta \phi$ will only be used to describe the linear free-field fluctuations.}. Its power spectrum can be related to the modulus-squared of the mode function for the field perturbation during inflation
\begin{equation}
\label{eq:power_spectrum}
\mathcal{P}_{\R} (k) = \frac{k^3}{2 \pi^2}\frac{ |\delta \phi_k|^2}{2 \epsilon_1} 
\, .
\end{equation}
We see from \Eqs{eq:homogeneous_delta_phi} and~\eqref{eq:power_spectrum} that $\mathcal{P}_{\R}$ remains constant in the long-wavelength limit when the decaying mode becomes negligible, corresponding to adiabatic perturbations~\cite{Gordon:2000hv, Romano:2015vxz, Jackson:2023obv}. More generally it is the amplitude of the growing mode on long wavelengths, $C$ in \eq{eq:homogeneous_delta_phi}, that determines the comoving curvature perturbation at the end of inflation which is then used to characterise the amplitude of primordial density fluctuations after inflation. 
Thus we are interested in determining the ``homogeneous'' part $\delta \phi_{k,\mathrm{h}}$ of the mode function $\delta \phi_k$ for each waveumber $k$ that survives in the long-wavelength limit as $k/aH\to0$.
If the behaviour of $\delta \phi_{k,\mathrm{h}}(N)$ is known at a given time $N_*$ (for example, at or soon after Hubble exit, $k=aH$), then the integration constants $C_k$ and $D_k$ can be assessed via continuous matching to the solution~\eqref{eq:homogeneous_delta_phi},
\begin{equation}
\label{eq:separate_universe_constants}
\begin{split}
    C_k & = \frac{\delta \phi_{k,\mathrm{h}*}}{\sqrt{2\epsilon_{1*}}} - a_{*}^3 H_{*} \sqrt{2\epsilon_{1*}} \left( \frac{\epsilon_{2*}}{2} \delta \phi_{k,\mathrm{h}*} - \partial_N \delta \phi_{k,\mathrm{h}*} \right) \int_{N_*} ^{N_{\mathrm{end}}}\frac{\dd\tilde{N}}{2 \tilde{\epsilon}_1 \tilde{a}^3 \tilde{H}} \, , \\
    D_k & = a_{*}^3 H_{*} \sqrt{2\epsilon_{1*}} \left( \frac{\epsilon_{2*}}{2} \delta \phi_{k,\mathrm{h}*} - \partial_N \delta \phi_{k,\mathrm{h}*} \right) .
\end{split}
\end{equation}
The linear power spectrum at the end of inflation is then
\begin{equation}
\label{eq:separate_universe_power_spectrum}
    \mathcal{P}_{\R} (k) = \frac{k^3}{2 \pi^2} |C_k|^2 \, .
\end{equation}
This relation, by construction, is exact when $C_k$ is the amplitude of the true growing mode. 
In Section~\ref{sec:noise_models} we will investigate different approaches to determine the amplitude of the homogeneous parts of linear field perturbations during inflation.

\subsection{Classical \texorpdfstring{$\delta \Nb$}{δN} formalism}

To study large perturbations originating from quantum field fluctuations during inflation, we will need to go beyond linear perturbation theory. One such method is the $\delta \Nb$ formalism~\cite{Starobinsky:1982ee, Starobinsky:1986fxa, Sasaki1996, Sasaki:1998ug, Wands:2000dp, Lyth:2004gb}. Here the curvature perturbation on uniform density hypersurface, is identified with the difference, $\delta \mathcal{N}$, between the local integrated expansion from an initial spatially-flat hypersurface to a final uniform-density hypersurface at the end of inflation, and the background expansion. In linear theory, we can identify this with the linear curvature perturbation at the end of inflation, $\R=H\delta\phi/\dot\phi=-\delta\Nb_{\mathrm{lin}}$, where the end of inflation hypersurface corresponds to a comoving (and hence uniform-field) hypersurface. Here, recall that $\R$ and $\delta\phi$ denote linear fluctuations, and we have introduced the number of \efolds until the end of inflation, $\mathcal{N}$.\footnote{The number of e-folds until the end of inflation is defined as
\begin{equation}
    \Nb(N) = \int_N^{N_{\rm end}} \d \tilde{N} \,,
\end{equation}
where we use a calligraphic font to distinguish this 
from the e-folding time, $N$, during inflation, since they go in opposite directions, $\d\Nb=-\d N$.} As we have seen, we can identify $\R$ with the amplitude, $C$, of the growing mode of the homogeneous solution~\eqref{eq:homogeneous_delta_phi} using the long-wavelength limit of linear perturbation theory.
More generally, we can calculate the local, non-linear expansion on large scales in the separate-universe approach using the local Klein--Gordon equation~\eqref{eq:klein_gordon} and local Friedmann equation~\eqref{eq:Friedmann}~\cite{Wands:2000dp,Lyth:2004gb}. 

In classical, single-field inflation the local expansion is determined by the local field value and its time derivative on an initial spatially-flat hypersurface at time $N_*$ at a point $\vec{x}$ in real space, which gives the final non-linear curvature perturbation on uniform-density hypersurfaces~\cite{Sasaki:1998ug,Malik:2008im}
\begin{equation}
\label{eq:delta_N_formalism}
\delta \Nb = \Nb\left(\phi_{*} + \delta\phi_{\mathrm{h}*}, \dot{\phi}_{*}+\dot{\delta \phi}_{\mathrm{h}*}\right)-\Nb\left(\phi_*,\dot{\phi}_*\right) \, .
\end{equation}
Here $\Nb(\phi, \dot{\phi})$ is the classical solution that gives the total number of \efolds from the phase-space position $(\phi, \dot{\phi})$ until the end of inflation. Hence we refer to this as the classical $\delta \Nb$ formalism. Since we are evaluating the expansion in the phase space for homogeneous cosmologies, we need to identify the homogeneous part, $\phi_{*} + \delta\phi_{\mathrm{h}*}$, of the full inhomogeneous field, $\phi_{*} + \delta\phi_{*}$, on the initial hypersurface, as well as its proper time derivative, $\dot{\phi}_{*}+\dot{\delta \phi}_{\mathrm{h}*}$.

The $\delta \Nb$ formalism is most often applied in cases where there is an attractor solution, reducing the 2D phase space to a 1D curve, $\dot{\phi} = \dot{\phi} (\phi)$.  Consider the local homogeneous field value, $\phi_{*} + \delta\phi_{\mathrm{h}*}$, at some time $N_*$ during inflation, on a spatially-flat hypersurface, which can be split into a background part and a perturbation.
As there is a one-to-one relation between $\phi$ and $N$ on the background trajectory, the local homogeneous field perturbation, $\delta\phi_{\mathrm{h}*}$, corresponds to a local adiabatic fluctuation, $\delta \Nb$, such that
\begin{equation}
\label{eq:perturbed_phi}
    \phi_{\mathrm{bg}} (N_*) + \delta \phi_{\mathrm{h}*} 
    =
    \phi_{\mathrm{bg}}(N_* - \delta \Nb) 
    \, .
\end{equation}
For monotonic $\phi_{\mathrm{bg}}(N)$ and $\delta\phi_{\mathrm{h}*}>0$ the sign of $\delta \Nb$ in \eq{eq:perturbed_phi} is positive if $\dot{\phi}<0$ and negative if $\dot{\phi}>0$. If we consider the field coarse-grained on some large scale, $k_{\mathrm{S}}<aH$, then in linear theory the perturbation, $\delta \phi$, is due to the sum of independent fluctuations on large scales $k_{\mathrm{in}}<k<k_{\mathrm{S}}$, where $k_{\mathrm{S}}$ is the smoothing scale. Thus $\delta \phi$ is the sum of many Gaussian variables and is itself a Gaussian, and the linear curvature perturbation, $\R = H\delta \phi/ \dot{\phi}$, is also a Gaussian. From the definition of the power spectrum, the coarse-grained curvature perturbation has a variance
\begin{equation}
\label{eq:variance_zeta_lin}
    \sigma_{\R}^2 (k_{\mathrm{S}}) = \int_{k_{\mathrm{in}}}^{k_{\mathrm{S}}} \mathcal{P}_{\R} (k) \d \ln k \, .
\end{equation}
Since $\R$ is constant on super-Hubble scales for adiabatic perturbations~\cite{Jackson:2023obv}, $\sigma_{\R}^2 (k_{\mathrm{S}})$ has no explicit $N_*$ dependence, and conservation of probability gives the probability density function (PDF) for $\delta\Nb$ as
\begin{equation}
\label{eq:delta_N_pdf_delta_phi}
    P(\delta\Nb) = \frac{e^{-\frac{\R^2}{2 \sigma_{\R}^2 (k_{\mathrm{S}})}}}{\sqrt{2 \pi} \sigma_{\R} (k_{\mathrm{S}})} \bigg| \frac{\d \R}{\d \delta\Nb} \bigg| \, .
\end{equation}
Therefore to apply the $\delta \Nb$ formalism, we need to know $\sigma_{\R} (k_{\mathrm{S}})$, the linear relation $\delta \phi = \dot{\phi}\R/H$, and the non-linear relation $\delta\Nb = \delta\Nb (\delta \phi)$ to determine $P(\delta\Nb)$.

Let us consider, as an example, a phase of constant roll, \ie constant $\epsilon_2$~\cite{Tomberg:2023kli}, which will play an important role in our analysis in Sec.~\ref{sec:stochastic_results}. Integrating \eq{eq:hubble_flow_parameters} for $i=1$ in this case gives
\begin{equation}
\label{eq:epsilon_1_constant_epsilon_2}
    \epsilon_1 = \epsilon_{1*} e^{\epsilon_2 (N - N_*)} \, .
\end{equation}
On the other hand, as mentioned below \eq{eq:hubble_flow_parameters}, $\epsilon_1=  \dot{\phi}^2 / (2 H^2) $, hence
\begin{equation}
\label{eq:phi_N_derivative}
    \frac{\d\phi}{\d N} = \pm \sqrt{2 \epsilon_1} \, .
\end{equation}
The choice of sign comes from taking the square root. We set it to match the dynamics of the background, and for explicitness let us consider the case where $\phi$ decreases with time classically, which corresponds to a negative sign. Integrating \eq{eq:phi_N_derivative} using \eq{eq:epsilon_1_constant_epsilon_2} gives the background solution
\begin{equation}
\label{eq:constant_epsilon_2_background_phi}
    \phi_{\mathrm{bg}} (N) = \phi_* + \frac{2}{\epsilon_2}\sqrt{2 \epsilon_{1*}} \left[1- e^{\frac{\epsilon_2}{2}(N- N_*)}\right] \, .
\end{equation}
From \Eqs{eq:perturbed_phi} and~\eqref{eq:constant_epsilon_2_background_phi}, one then obtains\footnote{A ``bg'' subscript has been omitted from $\epsilon_1$, and higher derivatives, for brevity, as there is no possibility for confusion with another variable here.}
\begin{equation}
\label{eq:constant_epsilon_2_R}
    \R = 
    -\frac{\delta\phi_{\mathrm{h}*}}{\sqrt{2\epsilon_{1*}}}
    = -\frac{2}{\epsilon_2} \left( 1- e^{-\frac{\epsilon_2}{2}\delta\Nb} \right) \, .
\end{equation}
We can now find $P(\delta\Nb)$ using \eq{eq:delta_N_pdf_delta_phi}, giving~\cite{Pi:2022ysn, Tomberg:2023kli, Inui:2024sce}
\begin{equation}
\label{eq:delta_N_pdf_constant_epsilon_2}
    P(\delta\Nb) = \frac{1}{\sqrt{2 \pi}  \sigma_{\R}(k_{\mathrm{S}})}\exp \left[ -\frac{2}{\sigma_{\R}^2 (k_{\mathrm{S}}) \epsilon_2^2} \left( 1- e^{-\frac{\epsilon_2}{2}\delta\Nb} \right)^2 - \frac{\epsilon_2}{2} \delta\Nb \right]  ,
\end{equation}
where $\sigma_{\R} (k_{\mathrm{S}})$ is given by \eq{eq:variance_zeta_lin}. The result has no explicit dependence on $N_*$, except through the value of $k_{\mathrm{S}}$.

By applying the classical $\delta \N$ formalism at a point $\vec{x}$, including fluctuations across a range of comoving scales, we can 
%use it to 
compare it against the stochastic formalism, the main interest of our investigation.

\subsection{Stochastic formalism}

Stochastic inflation is an effective-field-theory approach to modelling inflationary dynamics, where the full quantum-field-theory problem is mapped into a stochastic description. This is done by splitting the full quantum field, $\hat{\phi}$, and its momentum, $\hat{\pi}$, into short- and long-wavelength components, $\hat{\phi} = \phi + \hat{\phi}_\text{s}$ and $\hat{\pi} = \pi + \hat{\pi}_\text{s}$, where ``short'' and ``long'' are defined with respect to a \textit{coarse-graining} scale $k_{\sigma} = \sigma aH$. Short-wavelength modes originate as quantum vacuum fluctuations but they impart a stochastic kick to the long-wavelength field as they continuously cross the coarse-graining scale during inflation. Crucially, we assume that the Hubble damping drives the modes into a squeezed state on long wavelengths, which means that the quantum fluctuations can be treated as a classical random field~\cite{Polarski1996, Grain:2017dqa}. This allows the backreaction of perturbations to be included in the classical dynamics on long wavelengths. 
We will model the long-wavelength dynamics using the separate-universe approach, which allows us to use the same non-linear evolution equations as in an unperturbed, background cosmology, apart from the inclusion of stochastic noise.

The stochastic equations of motion for single-field inflation are 
\begin{equation}
\label{eq:stochstic_eom}
\begin{split}
    \frac{\partial \phi}{\partial N} & = \piphi  + \xi_{\phi} \,,
    \\
    \frac{\partial \piphi}{\partial N} & = - \left(3-\frac{\piphi^2}{2}\right)\left[ \piphi + \frac{1}{V(\phi)} \frac{\d V(\phi)}{\d \phi} \right]  + \xi_{\pi} \,,
    \end{split}
\end{equation}
where $\xi_{\phi}$ and $\xi_{\pi}$ are stochastic source terms, and $\phi$ and $\pi$ are now random variables. The covariance of the noise is identified with the quantum vacuum expectation value of the fields they stem from, which gives
\begin{equation}
\label{eq:noise_moments_simplified}
    \begin{split}
    \langle \xi_{\phi} ( N_1) \xi_{\phi} (N_2) \rangle & =  \frac{k_{\sigma}^3}{2\pi^2}  |\delta \phi_{k_\sigma,\mathrm{h}}(N_1)|^2 \delta_{\mathrm{D}} (N_1 - N_2) \, ,\\
    \langle \xi_{\pi} ( N_1)\xi_{\pi} ( N_2) \rangle & = \frac{k_{\sigma}^3}{2\pi^2} |\delta \pi_{k_\sigma,\mathrm{h}}(N_1)|^2 \delta_{\mathrm{D}} (N_1 - N_2) \, ,\\
    \langle \xi_{\phi} ( N_1) \xi_{\pi} ( N_2) \rangle & = \frac{k_{\sigma}^3}{2\pi^2} \Rea\left[ \delta \phi_{k_\sigma,\mathrm{h}}(N_1) \delta \pi_{k_\sigma,\mathrm{h}}^{*} (N_1) \right]  \delta_{\mathrm{D}} (N_1 - N_2) \, .
    \end{split}
\end{equation}
Here, we have considered the case where coarse graining is performed via a step function in Fourier space, with $\epsilon_1\ll 1$, and $\delta_{\mathrm{D}}$ is the Dirac distribution. A ``*'' superscript denotes the complex conjugate, and 
%only the real part of the cross product of mode functions is involved, since the imaginary part, which gives rise to the non-vanishing commutator of the field and its momentum at the quantum level, needs to be dropped in the stochastic description~\cite{Grain:2017dqa}. A
angle brackets $\langle \cdot \rangle$ are used to denote an ensemble average. Since the separate-universe approach is being used in the stochastic equations~\eqref{eq:stochstic_eom}, $\delta \phi_{k_\sigma,\mathrm{h}}$ and $\delta \pi_{k_\sigma,\mathrm{h}}$ in \eq{eq:noise_moments_simplified} correspond to the homogeneous part of the fluctuations~\cite{Jackson:2023obv}.

The stochastic system of equations~\eqref{eq:stochstic_eom} can be written in a matrix form. We introduce the covariance matrix
\begin{equation}
    \label{eq:covariance_matrix}
    \Xi
    = 
    \frac{k_{\sigma}^3}{2\pi^2}
    \begin{pmatrix}
    |\delta\phi_{k_\sigma,\mathrm{h}}|^2 & \Rea{ \left[\delta\phi_{k_\sigma,\mathrm{h}} \delta\pi_{k_\sigma,\mathrm{h}}^{*} \right]} \\
    \Rea{ \left[\delta\phi_{k_\sigma,\mathrm{h}} \delta\pi_{k_\sigma,\mathrm{h}}^{*} \right]} & |\delta\pi_{k_\sigma,\mathrm{h}}|^2
    \end{pmatrix} \, .
\end{equation}
 If $\Xi$ is non-singular, then its square-root, $S$, is real and symmetric. We can then rewrite \eq{eq:stochstic_eom} as
\begin{equation}
    \label{eq:full_langevin_equation}
    \frac{\d }{ \d N}
    \begin{pmatrix}
        \phi\\
        \pi_{\phi}
    \end{pmatrix}
    =
    \begin{pmatrix}
         \pi_{\phi}\\
        - \left(3-\frac{\pi^2}{2}\right)\left[ \pi_{\phi} + \frac{1}{V(\phi)} \frac{\d V(\phi)}{\d \phi} \right]
    \end{pmatrix}
    + 
    \begin{pmatrix}
    S_{\phi \phi} & S_{\phi \pi} \\
    S_{\phi \pi} & S_{\pi \pi}
    \end{pmatrix}
    \begin{pmatrix}
        {\xi}_{1}\\
        {\xi}_{2}
    \end{pmatrix} \, ,
\end{equation}
where $S^2 = \Xi$, with ${\xi}_1$ and ${\xi}_2$ being two independent, normally-distributed random variables with unit variance.

The covariance matrix~\eqref{eq:covariance_matrix} defines a quadratic form, the contours of which are ellipses in phase space. Their properties determine how far the stochastic system is from an effective one-dimensional noise. The area of the ellipses is related to the determinant of $\Xi$, which is proportional to the squared Wrońskian of the mode functions, which itself is set by the canonical commutator between the field and its momentum at the quantum level. Working with $\phi$ and $\pi_\phi$ as the phase-space coordinates, the area thus decays exponentially as $a^{-6}H^{-2}$. The eccentricity of the ellipses is related to the amount of quantum squeezing. As the mode gets more and more squeezed, the eccentricity approaches one. In this limit the ellipse collapses to its semi-major axis. The angle made between this semi-major axis and the $ \phi$ direction is given by
\begin{equation}
\label{eq:noise_phase_space_angle_2D}
    \theta_{\mathrm{n}} \equiv \arctan \left( \frac{\lambda_{+} - \Xi_{\phi \phi}}{\Xi_{\pi \phi}} \right)  ,
\end{equation}
where $\lambda_{+}$ is the positive eigenvalue of $\Xi$.

If $\delta \phi_{k_{\sigma}, \mathrm{h}}$ is dominated by a single mode in \eq{eq:homogeneous_delta_phi}, then the determinant of $\Xi$ exactly vanishes, and the ellipses are infinitely squeezed. There is then only one degree of freedom in the noise, as $\delta \phi_{k_{\sigma}, \mathrm{h}}$ and $\delta \pi_{k_{\sigma}, \mathrm{h}}$ are fully correlated. The 2D noise of the stochastic system reduces to 1D, although the phase space of the local field remains 2D in general. The angle of the noise then reduces to
\begin{equation}
\label{eq:noise_phase_space_angle_1D}
    \theta_{\mathrm{n}} \equiv  \arctan \left( \frac{\ \delta \pi_{k,\mathrm{h}}}{ \delta \phi_{k,\mathrm{h}} } \right)  .
\end{equation}
If $\tan\theta_{\mathrm{n}}=\tan\theta_{\mathrm{bg}}$ as well, where the background trajectory is at an angle $\tan\theta_{\mathrm{bg}} \equiv \partial_N \piphi/ \partial_N \phi = \epsilon_2/2$, then the noise is aligned with the background and the fluctuations are adiabatic. In this case the system given in \eq{eq:stochstic_eom} is fully described by a 1D Langevin system. 

The non-linear curvature perturbation~\eqref{eq:delta_N_formalism} can be found using the $\delta \Nb$ formalism in stochastic inflation~\cite{Vennin:2015hra}. Although $N$ is used as the time variable in~\eqref{eq:full_langevin_equation}, by solving the stochastic system of equations for the field, the local integrated expansion from an initial spatially-flat hypersurface to the final uniform-density hypersurface at the end of inflation, $\N$, becomes a stochastic variable. Finding $\N$ and its associated PDF, $P( \N)$, then becomes a first-passage-time problem~\cite{Vennin:2015hra}. The full non-linear curvature perturbation of a coarse-grained patch at the end of inflation is thus given by
\begin{equation}
    \delta \N = \N - \Nmean \, ,
\end{equation}
with an associated PDF, $P(\delta \N)$. While $P(\delta \N)$ can be found analytically in certain cases~\cite{Pattison:2017mbe, Ezquiaga:2019ftu, Pattison:2021oen, Animali:2022otk, Briaud:2023eae}, in general it needs to be found numerically. It has however been shown that in the presence of quantum diffusion we expect the far tail of $P(\delta \N)$ to be a simple exponential~\cite{Pattison:2017mbe}, potentially giving a significant enhancement of very large curvature fluctuations compared to the approximately Gaussian tail found in perturbation theory.

\section{Noise models}
\label{sec:noise_models}

\subsection{Consistency criterion}

Here we will detail how the homogeneous parts of the fluctuations on a given scale, $\delta \phi_{k_\sigma,\mathrm{h}}$ and $\delta \pi_{k_\sigma,\mathrm{h}}$ in \eq{eq:noise_moments_simplified}, can be calculated. There are a number of approaches one can take even in the context of linear perturbation theory, but a consistency requirement of any approach is that it gives the correct linear power spectrum at the end of inflation~\cite{Jackson:2023obv}. This is to make sure the correct constants of integration $C_k$ and $D_k$ have been chosen in the homogeneous solution~\eqref{eq:homogeneous_delta_phi} for each wavenumber $k$. Therefore we require that the fractional difference between the power spectrum at the end of inflation, \eq{eq:separate_universe_power_spectrum}, found by identifying the homogeneous part of the perturbations at $N_{\sigma}$, and the power spectrum~\eqref{eq:power_spectrum} found by numerically solving the full (inhomogeneous) mode equation~\eqref{eq:full_delta_phi}, is less than 1\%:
\begin{equation}
\label{eq:seperate_universe_consistancy_relation}
    \left| \frac{\mathcal{P}_{\R} (\delta \phi_{k_{\sigma},\mathrm{h}}, \partial_N \delta \phi_{k_{\sigma},\mathrm{h}},N_{\sigma}) - \mathcal{P}_{\R} (k, N_{\mathrm{end}})}{\mathcal{P}_{\R} (k, N_{\mathrm{end}})} \right| \leq 0.01 \, ,
\end{equation}
for all of the modes included in the numerical simulations. Here $N_{\sigma}$ is the matching time corresponding to $k = \sigma aH$. For the two potentials considered in this paper, the above criterion is shown in Appendix~\ref{app:noise_model_accuracy} for the noise model used in the simulations, as will be detailed below.

\subsection{Analytical approximations}
\label{sub:noise_analytical}

The simplest approach, commonly adopted in the $\delta N$ formalism~\cite{Vennin:2015hra, Pattison:2017mbe, Biagetti:2018pjj, Ezquiaga:2019ftu, Firouzjahi:2018vet}, is to approximate the noise as that for a light field in a quasi-de Sitter background\footnote{In the de Sitter limit, $\epsilon_1\to 0$, the field perturbations decouple from the metric and we effectively have a test field in de Sitter spacetime.}, corresponding to slow-roll inflation with $|\epsilon_i| \ll 1 \ \forall i\ \geq 1$. The general solution for the homogeneous part of the field perturbation in the spatially-flat gauge can then be readily identified as~\cite{Jackson:2023obv}
\begin{equation}
    \label{eq:delta_phi_de_sitter_homogeneous}
    \delta \phi_{k, \mathrm{h}} = \frac{iH}{\sqrt{2k^3}} \left[ \left(\alpha_k-\beta_k \right) -i \left( \alpha_k + \beta_k \right) \frac{(k\eta)^3}{3}  \right] .
\end{equation}
On super-Hubble scales the decaying mode, ${\sim (k \eta)^3}$, exponentially decays, and is dropped. In particular, if Bunch--Davies initial conditions ($\alpha_k=1$ and $\beta_k=0$) are used, then the covariance matrix~\eqref{eq:covariance_matrix} is
\begin{equation}
    \label{eq:covariance_de_sitter}
    \Xi = \frac{H^2}{(2\pi)^2}
    \begin{pmatrix}
        1 & 0 \\
        0 & 0
    \end{pmatrix} \, .
\end{equation}
Note that in this simple limit the homogeneous field perturbation is constant and hence the noise is independent of the choice of coarse-graining scale, $k_\sigma$. 
As the noise is only in the $\phi$ direction, only a single random variable is needed to characterise the noise at each time step. The consistency relation~\eqref{eq:seperate_universe_consistancy_relation} is obeyed for slow-roll models with $|\epsilon_2| \ll 1 $.

This approach can be further refined by requiring $\epsilon_1 \ll 1$  but allowing $|\epsilon_i| \neq 0 \ \forall i {\geq 2}$. We will require
\begin{equation}
\label{eq:nu_squared}
    \nu^2 = \frac{9}{4} - \epsilon_1 + \frac{3}{2} \epsilon_2 -  \frac{1}{2} \epsilon_1\epsilon_2 + \frac{1}{4} \epsilon_2^2 + \frac{1}{2} \epsilon_2\epsilon_3
\end{equation}
to be approximately constant. This gives an analytic approximation for the general solution of the mode equation \eq{eq:full_delta_phi} of the form~\cite{Stewart:1993bc}
\begin{equation}
\label{eq:delta_phi_bessel}
    \delta \phi_k = \frac{\sqrt{-\eta} }{a}[A_k J_{\nu}(-k\eta) + B_k Y_{\nu}(-k\eta)] \, ,
\end{equation}
where $J_{\nu}$ and $Y_{\nu}$ are Bessel functions of the first and second kind. If we again assume Bunch--Davies initial conditions~\eqref{eq:bunch_davies}, then the Bessel coefficients are given by
\begin{equation}
\label{eq:A_k_and_B_k}
A_k = \frac{\sqrt{\pi}}{2}e^{i \left( \nu + \frac{1}{2} \right) \frac{\pi}{2}} \quad , \quad B_k = \frac{\sqrt{\pi}}{2}e^{i \left( \nu + \frac{3}{2} \right) \frac{\pi}{2}} \, .
\end{equation}
The homogeneous component is defined as the part of \eq{eq:delta_phi_bessel} which is a solution of \eq{eq:seperate_universe_approach_delta_phi} and is found by taking the leading order term in a $(-k\eta)$ expansion of $J_{\nu}$ and $Y_{\nu}$. The homogeneous growing mode is thus given by
\begin{equation}
    \delta \phi_{k,\mathrm{h}} = e^{i \left( \nu -\frac{1}{2} \right)\frac{\pi}{2}} \frac{H}{2^{\frac{3}{2}-\nu}} \frac{(-k \eta)^{\frac{3}{2} - \nu}}{\sqrt{2 k^3}} \, .
\end{equation}
The associated correlation matrix~\eqref{eq:covariance_matrix}, which is evaluated at $-k\eta = \sigma$, is then
\begin{equation}
    \label{eq:covariance_constant_nu}
    \Xi = \frac{H^2}{4\pi^2} \left[ \frac{\Gamma (\nu)}{\Gamma (\frac{3}{2})} \right]^2 \left(\frac{\sigma}{2}\right)^{3-2\nu}
    \begin{pmatrix}
        1 &  \nu - \frac{3}{2} \\
        \nu - \frac{3}{2} & \left( \nu - \frac{3}{2} \right)^2
    \end{pmatrix} \, .
\end{equation}
While there is noise in both the $\phi$ and $\piphi$ directions, the determinant of this matrix is exactly zero, hence there is only one degree of freedom and the stochastic system is described by a single (1D) noise term, as was discussed near \eq{eq:noise_phase_space_angle_1D}. One can verify that the consistency relation~\eqref{eq:seperate_universe_consistancy_relation} is satisfied for slow-roll inflation, but we will also show that the consistency relation is satisfied in non-slow-roll models after a sudden transition, including, but not limited to, constant-roll regimes. Note that for $\nu\to3/2$, we recover the result~\eqref{eq:covariance_de_sitter} for a massless field in de Sitter and the $\sigma$ dependence is lost.

\subsection{Numerical matching}
\label{sec:Numerical:Matching}

The analytic approximations used to estimate the noise in \Sec{sub:noise_analytical} rely on $\nu^2$ in \eq{eq:nu_squared} being constant. This will not be valid when the inflaton experiences a sudden transition. Note that, during sudden transitions, not only scales around and above the Hubble radius are affected, sub-Hubble modes are also altered, and this implies that when they later cross the Hubble radius they may do so with ``initial'' conditions that depart from the Bunch--Davies profile. 

The most general approach to determine the noise is to match the general homogeneous solution~\eqref{eq:homogeneous_delta_phi} to the full numerical solution for a given scale, $\delta \phi_k$ and $\partial_N \delta \phi_k$, on sufficiently long wavelengths. The covariance matrix is then
\begin{equation}
    \label{eq:covariance_matrix_direct}
    \Xi
    = 
    \frac{k_{\sigma}^3}{2\pi^2}
    \begin{pmatrix}
    |\delta\phi_{k_\sigma}|^2 & \Re{ \left[\delta\phi_{k_\sigma } \delta\pi_{k_\sigma }^{*} \right]} \\
    \Re{ \left[\delta\phi_{k_\sigma } \delta\pi_{k_\sigma}^{*} \right]} & |\delta\pi_{k_\sigma }|^2
    \end{pmatrix} \Bigg|_{N=N_{\sigma}} \, ,
\end{equation}
with $\sigma$ sufficiently small such that the behaviour of $\delta \phi_{k_{\sigma}}$ has minimal contribution from gradient, ${\cal O}(k^2)$ terms. This approach can be used for any background dynamics and can account for non-Bunch--Davies initial conditions but is limited to scales significantly larger than the Hubble scale, $\sigma\ll1$. As \eq{eq:covariance_matrix_direct} has a non-zero determinant, the associated phase space is 2D and two random variables are required to describe the system.

While general, this approach must either accept a finite $k^2$ contamination or use a very small $\sigma$~\cite{De:2020hdo, Figueroa:2020jkf, Figueroa:2021zah, Ahmadi:2022lsm, Mishra:2023lhe, Ballesteros:2024pwn}, potentially missing non-linear effects associated with stochastic inflation~\cite{Jackson:2023obv}. We find that the consistency relation~\eqref{eq:seperate_universe_consistancy_relation} is obeyed, with $C_k$ found using \eq{eq:separate_universe_constants} with $\delta \phi_{k, \mathrm{h}*}$ and $\delta \pi_{k, \mathrm{h}*}$ replaced by $\delta \phi_{k*}$ and $\delta \pi_{k*}$, when $\sigma \leq 0.01$ in the absence of a sudden transition. An even smaller value is required to be accurate for all modes through a sudden transition.

\subsection{Semi-analytical approach}
\label{sub:semi_analytical}

The third approach is to try and combine the advantages of the previous two. Namely being able to directly identify the homogeneous component with the analytical approach, and allowing for non-Bunch--Davies initial conditions with the numerical approach, effectively using the Bessel function~\eqref{eq:delta_phi_bessel} as an ansatz.

This is done by assuming that $\nu$ is locally constant in \eq{eq:full_delta_phi}, and continuously matching the numerical solution to \eq{eq:delta_phi_bessel} to find the Bessel coefficients
\begin{equation}
\label{eq:bessel_matching}
\begin{split}
    A_k &= \frac{a}{\sqrt{-\eta}}\frac{ \delta \phi_{k} \partial_N \left[ Y_{\nu}(-k\eta)\right]  -  \left( \frac{3}{2}\delta \phi_k - \partial_N \delta \phi_k  \right) Y_{\nu}(-k\eta)}{J_{\nu}(-k\eta)\partial_N [Y_{\nu}(-k\eta)] -  \partial_N[J_{\nu}(-k\eta)] Y_{\nu}(-k\eta)} \bigg|_{N=N_{\sigma}}\, , \\
    B_k &= -\frac{a}{\sqrt{-\eta}}\frac{ \delta \phi_{k} \partial_N \left[ J_{\nu}(-k\eta)\right]  -  \left( \frac{3}{2}\delta \phi_k - \partial_N \delta \phi_k  \right) J_{\nu}(-k\eta)}{J_{\nu}(-k\eta)\partial_N [Y_{\nu}(-k\eta)] -  \partial_N[J_{\nu}(-k\eta)] Y_{\nu}(-k\eta)} \bigg|_{N=N_{\sigma}}\, .
\end{split}
\end{equation}
This method allows for possible time dependence of $\nu$ before Hubble exit, hence for non-trivial sub-Hubble dynamics during the transition. 

The homogeneous behaviour can then be identified as before, using the leading terms in a $(-k\eta)$ expansion of \eq{eq:delta_phi_bessel}. One then finds
\begin{equation}
\label{eq:bessel_matched_homogeneous}
\delta \phi_{k,\mathrm{h}} =  \frac{\sqrt{-\eta}}{a} \left[ - \frac{B_k}{\pi} \Gamma \left( \nu \right)  + \frac{A_k \tan \left(\pi \nu \right) + B_k }{\Gamma \left( \nu +1 \right) \tan \left(\pi \nu \right)}  \left(- \frac{k\eta}{2} \right)^{2\nu} \right] \left(- \frac{k\eta}{2} \right)^{-\nu} \, ,
\end{equation}
where the leading term is the homogeneous growing mode and the second term is the homogeneous decaying mode, as $\eta \rightarrow 0$ at late times and $\nu>0$. Here $\epsilon_1 \ll 1$ was assumed, and $\nu$ is evaluated at the time $k$ crosses the coarse-graining scale (hence it carries an implicit $k$ dependence). If the growing mode dominates over the decaying mode, the covariance matrix is then given by
\begin{equation}
    \label{eq:covariance_constant_nu_bessel_matched}
    \Xi = \frac{H^2}{4\pi^2} \frac{4|B_k|^2}{\pi}  \left[ \frac{\Gamma (\nu)}{\Gamma (\frac{3}{2})} \right]^2 \left(\frac{\sigma}{2}\right)^{3-2\nu}
    \begin{pmatrix}
        1 &  \nu - \frac{3}{2} \\
        \nu - \frac{3}{2}& \left( \nu - \frac{3}{2} \right)^2
    \end{pmatrix} \, .
\end{equation}
When $B_k$ is given by its Bunch--Davies value~\eqref{eq:A_k_and_B_k}, then the above coincides with \eq{eq:covariance_constant_nu}. In this limit the determinant of the covariant matrix vanishes, and the stochastic system is subject to a 1D noise term [see again the discussion near \eq{eq:noise_phase_space_angle_1D}].

The homogeneous growing mode found in \eq{eq:covariance_constant_nu_bessel_matched} obeys the consistency relation~\eqref{eq:seperate_universe_consistancy_relation} as long as $\nu$ is approximately constant as the coarse-graining scale is crossed, see \App{app:noise_model_accuracy} for more details.
In practice, we define the period where the constant-$\nu$ approximation is valid by requiring that the consistency relation~\eqref{eq:seperate_universe_consistancy_relation} holds.

\begin{figure}
\begin{center}
        \includegraphics[width=\figurewidth\textwidth]{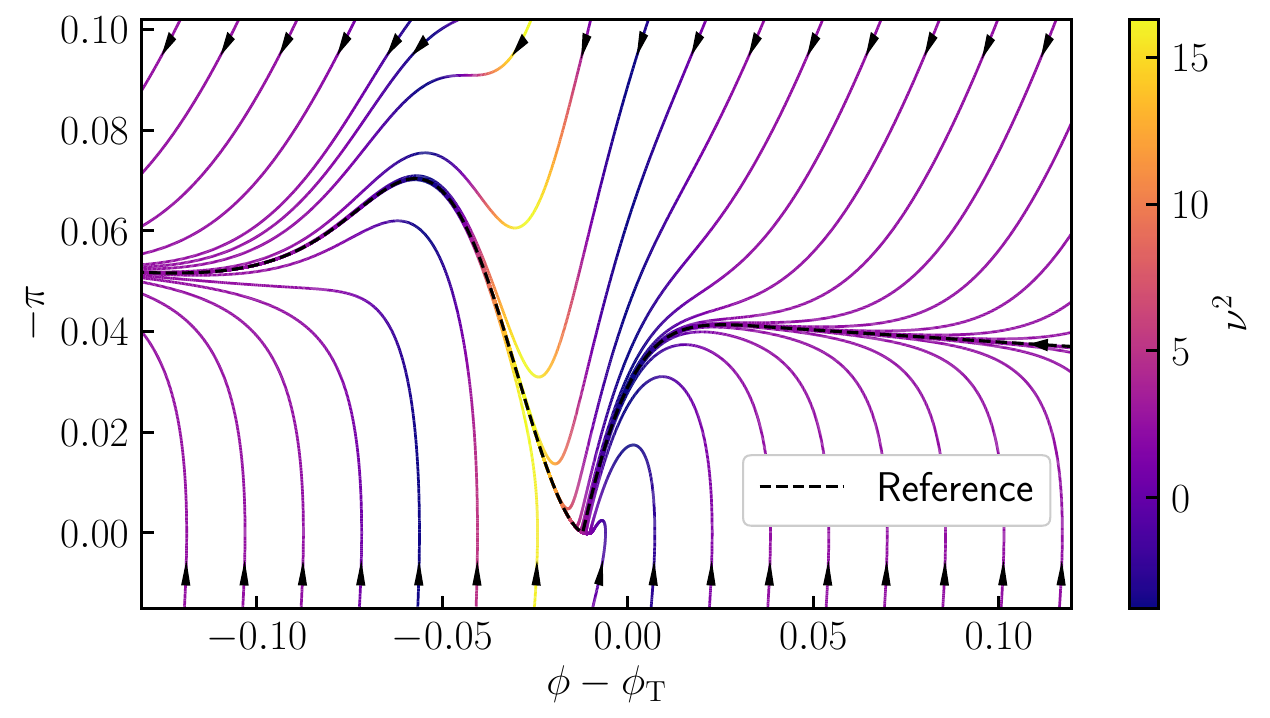}
        \caption{The phase-space behaviour for the Gaussian bump potential~\eqref{eq:potential_swagat}, for field values near the sudden transition. Here $\phi_{\mathrm{T}}$ is the field value at the first time when $\epsilon_2<-3$ on the reference trajectory which starts on the slow-roll attractor at early times. The colour bar shows the value of $\nu^2$ defined in \eq{eq:nu_squared}, at each point in the phase space.}
        \label{fig:swagat_phase_space}
\end{center}
\end{figure}

We expect a phase of approximately constant $\nu$ after a sudden transition for most singe-field models of inflation that produce a measurable abundance of PBHs. This can be seen by using the Klein--Gordon equation~\eqref{eq:klein_gordon} and the Hubble-flow parameters~\eqref{eq:hubble_flow_parameters} to rewrite \eq{eq:nu_squared} as
\begin{equation}
    \label{eq:nu_squared_with_epsilon_1}
    \nu^2 =  \frac{9}{4} - \epsilon_1 -\frac{1}{2} \frac{\epsilon_1 \epsilon_2^2}{(3 - \epsilon_1)}  + \left( \epsilon_1- 3 \right)\left[\frac{1}{V} \frac{\d^2 V}{\d \phi^2}  - \left(\frac{1}{V} \frac{\d V}{\d \phi} \right)^2 \right] \, .
\end{equation}
Note that this expression is exact. Post-transition, to grow the power spectrum sufficiently to produce a measurable abundance of PBHs, we have $\epsilon_1 \ll 1$. Then taking \eq{eq:nu_squared_with_epsilon_1} to zeroth-order in $\epsilon_1$, while requiring\footnote{In single-field models of interest, the maximum value of $|\epsilon_2|$ obtainable is $\mathcal{O} (10)$. Therefore $\epsilon_1 \ll 10^{-2}$ is required for $\epsilon_1 \epsilon_2^2 \ll 1$, which is easily obtained for most PBH producing models.} $\epsilon_1 \epsilon_2^2 \ll 1$, one obtains
\begin{equation}
\label{eq:nu_squared_approximation}
    \nu^2 \simeq \frac{9}{4} - 3\left[\frac{1}{V} \frac{\d^2 V}{\d \phi^2}  - \left(\frac{1}{V} \frac{\d V}{\d \phi} \right)^2 \right] \, .
\end{equation}
Therefore post-transition $\nu^2$ is only a function of $\phi$. This can be checked in \fig{fig:swagat_phase_space}, where the colour gradient encoding variations of $\nu^2$ is aligned with the $\phi$ direction, regardless of the phase-space trajectory. In fact, the maximum difference between \eq{eq:nu_squared_approximation} and $\nu^2$ for any of the trajectories shown is 0.05.

Taking the $N$ derivative of \eq{eq:nu_squared_approximation} then gives
\begin{equation}
\label{eq:nu_squared_approximation_N_derivative}
    |\partial_N \nu^2| \simeq  \sqrt{18 \epsilon_1} \left|\frac{1}{V} \frac{\d^3 V}{\d \phi^3} -  \frac{3}{V^2}\frac{\d V}{\d \phi} \frac{\d^2 V}{\d \phi^2}  + \left(\frac{1}{V} \frac{\d V}{\d \phi} \right)^3 \right| \, .
\end{equation}
In the limit where $\epsilon_1 \ll 1$, we can take $ \sqrt{18\epsilon_1} < 1$. In the models we will consider, apart from at a sudden transition, the potential derivative terms in the square brackets are at a maximum of order unity. Therefore in the post-transition phase, $|\partial_N \nu^2| < 1$ and $\nu$ is then approximately constant. The duration of this constant $\nu$ phase depends on the potential and the minimum value of $\epsilon_1$ reached.

\section{Stochastic \texorpdfstring{$\delta \Nb$}{δN} simulations and results}

\label{sec:stochastic_results}

In this section we will apply the techniques developed in \Sec{sec:noise_models} to describe the quantum noise used to calculate the full probability distribution for the first-passage time in the stochastic $\delta \Nb$ formalism. We compare this to predictions using the classical $\delta \Nb$ approach in models involving a sudden transition from an initial slow-roll inflation attractor, through an intermediate ultra-slow-roll phase (where $\epsilon_2 \sim -6$), before returning to an attractor solution at late times.

\subsection{Numerical approach}
\label{sub:numerical_approach}

We simulate the Langevin system given in \eq{eq:full_langevin_equation} using Euler--Maruyama steps~\cite{Kloeden1992}. The $(i+1)^\text{th}$ step is given by
\begin{equation}
\begin{split}
\label{eq:euler_step_phi_and_pi}
\phi^{(i+1)} & = \phi^{(i)} + \pi ^{(i)} \Delta N + \left( S_{\phi \phi}^{(i)} \xi_1^{(i)} + S_{\phi \pi}^{(i)} \xi_2^{(i)} \right) \sqrt{\Delta N} \, , \\
\pi^{(i+1)} & = \pi^{(i)} -  \left[3-\frac{\left(\pi^{(i)}\right)^2}{2}\right]\left[ \pi^{(i)} + \frac{1}{V(\phi^{(i)})} \frac{\d V(\phi^{(i)})}{\d \phi} \right]  \Delta N +\\
 & \quad \left( S_{\pi \phi}^{(i)} \xi_1^{(i)} + S_{\pi \pi}^{(i)} \xi_2^{(i)} \right) \sqrt{\Delta N} \, ,
\end{split}
\end{equation}
where $\xi_1^{(i)}$ and $\xi_2^{(i)}$ are independent random numbers drawn from a Gaussian distribution and $\Delta N$ is the time step. The system is initiated at ($\phi_{\mathrm{init}}$, $\pi_{\mathrm{init}}$). It is then propagated using the above Euler--Maruyama steps until $\epsilon_1(\phi^{(n)},\pi^{(n)})$ crosses one and the first-passage time $\N = n \Delta N$ is recorded, where $n$ is the number of steps required. By running many such simulation runs, $P(\delta \N)$ can be sampled numerically.

In principle, the noise covariance matrix, $\Xi$ in \eq{eq:covariance_matrix}, for each simulation should be calculated for the quantum field within that specific stochastic realisation. This can be done in two ways: either using the exact numerical-matching approach presented in \Sec{sec:Numerical:Matching}, or using the semi-analytical one discussed in \Sec{sub:semi_analytical}. While the former greatly increases the computational complexity, requiring supercomputers to solve with current techniques~\cite{Figueroa:2020jkf, Figueroa:2021zah}, the latter is more straightforward to implement since it only requires to compute the local value of $\nu$ in the background at each time step (the main limitation being that $\nu$ needs to be approximately constant). Although one of these methods would be required for a detailed investigation of the far tail, in what follows we rather evaluate the noise using the unperturbed, classical background. The reason is that, since the main goal of this work is to design a robust modeling of the noise, and to generalise importance-sampling methods to multi-dimensional phase-spaces, it is important for us to be able to compare our results with the classical-$\delta \Nb$ formalism in order to test their validity. This is why we solve stochastic equations in a setup that is otherwise classical: the noise is pre-computed along a reference, classical trajectory\footnote{In Eqs.~\eqref{eq:stochstic_eom}--\eqref{eq:full_langevin_equation} the \textit{local} time in terms of the number of \efolds, $N$, is used throughout. However, by limiting ourselves to using the pre-computed noise, in practice our simulations are run with the \textit{global} time $N$.} and after a fixed number of time steps, corresponding to a chosen range of comoving wavenumbers, the noise is turned off and the time to reach the end of inflation classically is recorded. With this approach, each simulation set presented used $10^5$ numerical runs, taking less than $15$ minutes on an 8 core 1.6 GHz processor. For more details about the computation and open source code, see the \href{https://github.com/Jacks0nJ/PyFPT}{\pyfpt} Github repository. The main speedup comes from the technique of importance sampling~\cite{Jackson:2022unc}, which we have extended to the 2D phase space.

To probe the rare and large $\delta \N$ events associated with PBH production efficiently, importance sampling is utilised~\Refa{Jackson:2022unc}.
Importance sampling introduces a bias term, $\mathcal{B}$, at each numerical step in \eq{eq:euler_step_phi_and_pi} to increase the frequency of large first-passage-time events being realised. The induced change to the relative probability for each step is then recorded, with the total relative probability of the biased path compared to the unbiased path being given by the weight, $w$. For an unbiased multi-dimensional Langevin equation, the $(i+1)^\text{th}$ step is given by
\begin{equation}
\vec{x}^{(i+1)} - \vec{x}^{(i)} = \vec{D}(\vec{x}^{(i)}, N^{(i)}) \Delta N + S(\vec{x}^{(i)}, N^{(i)}) \vec{\hat{\xi}}^{(i)} \sqrt{\Delta N} \, , 
\end{equation}
with drift vector $\vec{D}$ and noise matrix $S$. The biased step is
\begin{equation}
\vec{x}^{(i+1)} - \vec{x}^{(i)} = \vec{D}(\vec{x}^{(i)}, N^{(i)}) \Delta N + \vec{\mathcal{B}}(\vec{x}^{(i)}, N^{(i)}) \Delta N + S(\vec{x}^{(i)}, N^{(i)}) \vec{\hat{\xi}}^{(i)} \sqrt{\Delta N} \, .
\end{equation}
The weight of the whole stochastic process for $n$ steps is
\begin{equation}
\label{eq:weight}
w  = \exp \left( - \sum_{i=0}^{n-1} \Delta A^{(i)} \right) \, ,
\end{equation}
where~\cite{Mazonka:1998ge}
\begin{equation}
\label{eq:weight_A}
\Delta A^{(i)} = \left(\frac{\vec{\mathcal{B}}^{(i)}}{2} \Delta N + S^{(i)} \vec{\hat{\xi}}^{(i)} \sqrt{\Delta N} \right)^{\mathrm{T}}\left( \Xi^{(i)} \right)^{-1}\vec{\mathcal{B}}^{(i)} \, ,
\end{equation}
and $\Xi = S^2$ by definition. \pyfpt is able to apply importance sampling to a 2D Langevin system, as long as the correlation matrix is non-singular.

The error in the reconstruction of $P(\delta \N )$ is found numerically. When direct sampling is used, corresponding to $\vec{\mathcal{B}} = 0$, as each simulation run is independent, jackknife resampling is used. When importance sampling is used, the error in $P(\delta \N )$ is found from the error in the estimate of the mean of a lognormal distribution. See Ref~\cite{Jackson:2022unc} for details of both methods.

As already mentioned, when the growing mode dominates over the decaying mode, the noise becomes effectively 1D. In \App{app:noise_model_accuracy} we check that this a valid approximation even at Hubble crossing with $\sigma=1$, as long as $\nu$ is approximately constant. Since this condition is also required for the semi-analytical noise description of \Sec{sub:semi_analytical} to apply, we will focus on this regime from now on. In practice, this means that we will start our simulations \textit{after} a sudden transition to ultra-slow roll, once $\nu^2$ is approximately constant, although \pyfpt is applicable beyond this so long as the separate-universe approach applies.  

The implementation of importance sampling is more straightforward in 1D. If the noise is given solely by a single growing mode, then the system in general can be written as
\begin{equation}
\label{eq:stochastic_eom_1D}
\begin{split}
    \frac{\d \phi}{\d N} & = \pi + S_{\phi \phi} \hat\xi \, , \\
    \frac{\d \pi}{\d N} & = -\left(3- \frac{\pi^2}{2} \right) \left[\pi + \frac{1}{V(\phi)} \frac{\d V(\phi)}{\d \phi} \right] + \tan \left( \theta_{\mathrm{n}} \right) S_{\phi \phi} \hat\xi \, ,
\end{split}
\end{equation}
where $\hat\xi$ is a random variable with unit variance and $\tan (\theta_{\mathrm{n}})$ is given in \eq{eq:noise_phase_space_angle_1D}. Therefore the system is 2D but with a 1D noise. If $\tan (\theta_{\mathrm{n}}) =\partial_N \pi/ \partial_N \phi  \equiv \epsilon_2/2$ throughout, then the noise is aligned with the unperturbed background trajectory and the phase space is also 1D, \eg during slow roll or constant roll. However this is not in general the case in the absence of a dynamical attractor, \eg during ultra-slow roll, and in this case the full 2D phase space needs to be used to account for stochastic kicks misaligned with the unperturbed background trajectory.

Applying a bias to \eq{eq:stochastic_eom_1D} gives for the $(i+1)^\text{th}$ step
\begin{equation}
\label{eq:importance_sampling_1D_noise_step}
\begin{split}
    \Delta \phi^{(i+1)} & = \pi^{(i)} \Delta N + \mathcal{B}^{(i)}_{\phi} \Delta N + S^{(i)}_{\phi \phi} \hat{\xi} ^{(i)}\sqrt{\Delta N} \, ,   \\
    \Delta \pi^{(i+1)} & = \left(3- \frac{(\pi^{(i)})^2}{2} \right) \left[\pi^{(i)} + \frac{1}{V(\phi^{(i)})} \frac{\d V(\phi^{(i)})}{\d \phi} \right] \Delta N +  \mathcal{B}^{(i)}_{\pi} \Delta N +  \\
    & \quad \tan \left( \theta_{\mathrm{n}}^{(i)} \right) S^{(i)}_{\phi \phi} \hat{\xi}^{(i)} \sqrt{\Delta N}  \, .
\end{split}
\end{equation}
As there is only a single noise term, \eq{eq:weight_A} does not apply directly, as the determinant of $\Xi = S^2$ is zero and it is thus singular. Instead, the relative probability of the importance sampling step must be the same along each dimension. Therefore we apply the 1D version of \eq{eq:weight_A} for both $\phi$ and $\pi$ and equate them, giving the following constraint 
\begin{multline}
    \frac{\mathcal{B}^{(i)}_{\phi}}{(S^{(i)}_{\phi \phi})^2} \left( \frac{\mathcal{B}^{(i)}_{\phi} }{2}\Delta N +  S^{(i)}_{\phi\phi} \xi^{(i)} \sqrt{\Delta N} \right) = \\
    \frac{\mathcal{B}^{(i)}_{\pi}}{\left[\tan \left( \theta_{\mathrm{n}}^{(i)} \right) S^{(i)}_{\phi \phi} \right]^2} \left[ \frac{\mathcal{B}^{(i)}_{\pi} }{2}\Delta N +  \tan \left( \theta_{\mathrm{n}}^{(i)} \right) S^{(i)}_{\phi\phi} \xi^{(i)} \sqrt{\Delta N} \right]  ,
\end{multline}
which reduces to
\begin{equation}
    \mathcal{B}^{(i)}_{\phi} = \mathcal{B}^{(i)}_{\pi} \tan \left( \theta_{\mathrm{n}}^{(i)} \right)  .
\end{equation}
For diffusion-based bias with $\mathcal{B}^{(i)}_{\phi} = \mathcal{A} S^{(i)}_{\phi \phi}$, where $\mathcal{A}$ is known as the bias amplitude, one then finds
\begin{equation}
\label{eq:2D_1D_noise_diussion_based_bias}
    \mathcal{B}^{(i)}_{\phi} = \mathcal{A} \tan \left( \theta_{\mathrm{n}}^{(i)} \right) S^{(i)}_{\phi \phi} \, .
\end{equation}
It is this form of the bias that will be used in the numerical results below.

\subsection{Piece-wise linear potential}
\label{sub:starobinsky}

\begin{figure}
\begin{center}
        \includegraphics[width=0.48\textwidth]{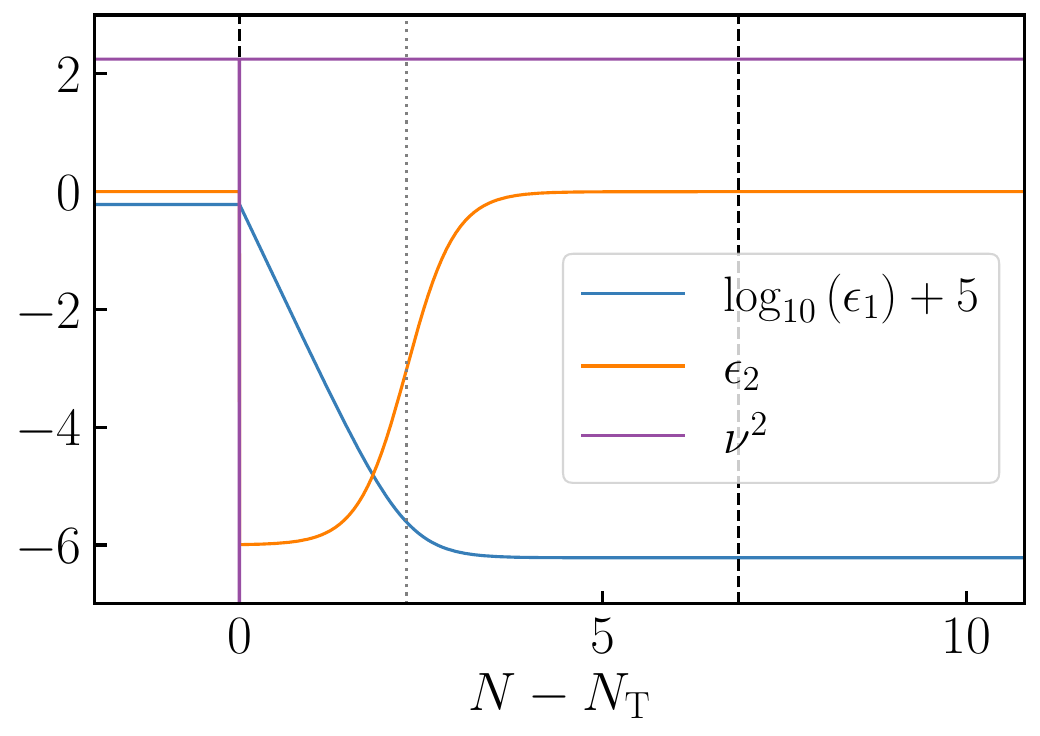}
        \includegraphics[width=0.51\textwidth]{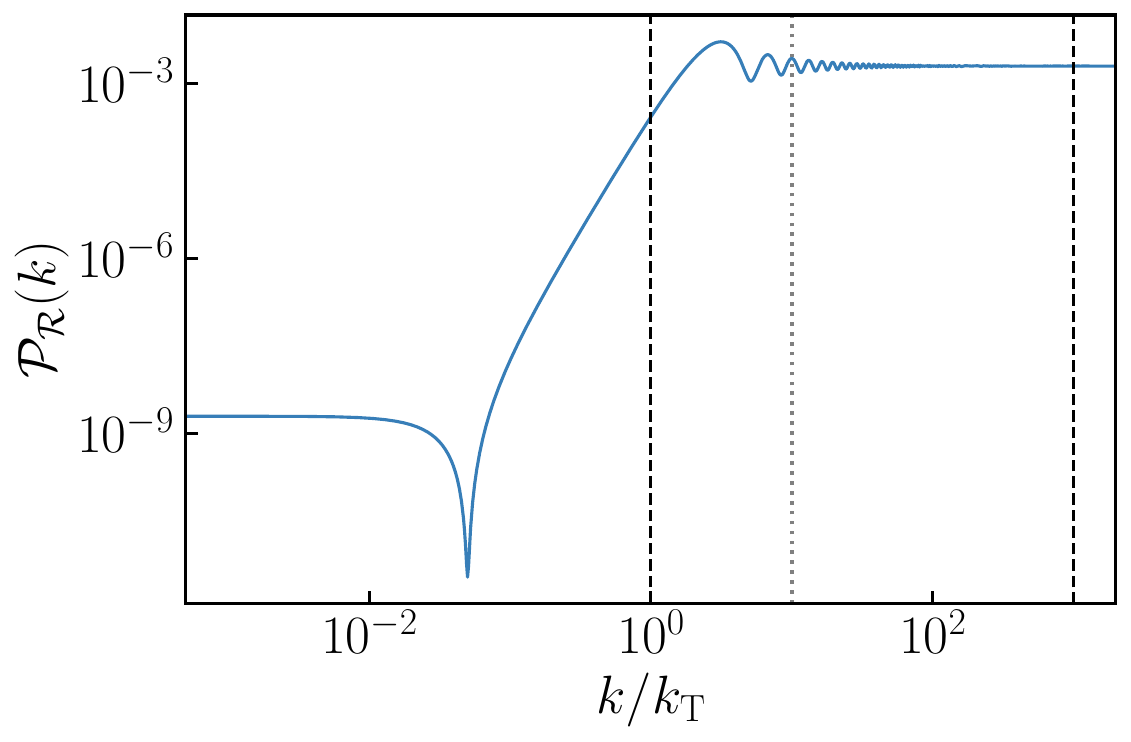}
        \caption{Left: the Hubble-flow parameters~\eqref{eq:hubble_flow_parameters} and $\nu^2$~\eqref{eq:nu_squared} for the piece-wise linear potential defined in \eq{eq:potential_starobinsky}, near the sudden transition at $N_{\mathrm{T}}$. The quantity $\log_{10}(\epsilon_1)+5$ is plotted to fit on the same vertical scale as $\epsilon_2$. Right: the associated power spectrum~\eqref{eq:power_spectrum} where $k_{\mathrm{T}}$ is the scale that crosses the Hubble radius at time $N_{\mathrm{T}}$.
 The gray dotted vertical line corresponds to when ultra-slow roll ends, here defined as when $\epsilon_2$ crosses $-3$, and the black dashed lines bound the modes used in our stochastic simulations, $k_{\mathrm{min}}<k<k_{\mathrm{max}}$ (the vertical lines on the left panel label the times at which the scales labeled by the same lines on the right panel cross the Hubble radius).} 
        \label{fig:starobinsky_set_up}
\end{center}
\end{figure}

Let us first consider one of the simplest models with a sudden transition from slow-roll to ultra-slow-roll inflation as a test case for \pyfpt, the piece-wise linear potential~\cite{Starobinsky:1992ts}. This toy model has many well-known analytical results~\cite{Leach:2001zf, Martin:2011sn, Martin:2014kja, Ahmadi:2022lsm, Pi:2022zxs}. The potential is 
\begin{equation}
    \label{eq:potential_starobinsky}
    V (\phi) = \begin{cases}
    V_0 + A_+ (\phi - \phi_{\mathrm{T}}) & \text{for } \phi\geq \phi_{\mathrm{T}}\, ,\\ 
    V_0 + A_- (\phi - \phi_{\mathrm{T}})& \text{for } \phi < \phi_{\mathrm{T}} \, ,
\end{cases}
\end{equation}
where $A_+>A_->0$ and we choose $\phi_{\mathrm{init}} \gg \phi_{\mathrm{T}}$ such that the slow-roll attractor, $\dot\phi=-A_+/3H$, has been reached before the transition. Immediately after the transition, there is a mismatch between the scalar-field velocity and the potential gradient, $A_-$, resulting in slow roll being broken and giving a transitory phase of ultra-slow roll (as shown in the left-hand panel of \fig{fig:starobinsky_set_up}). Eventually the field relaxes back onto a new slow-roll trajectory with $\dot\phi=-A_-/3H$. In the following we take $V_0 = 2.8719 \times 10^{-12}$ and $A_+ = 10^{-14}$ to have a scale-invariant power spectrum with an amplitude in agreement with CMB observations on large scales~\cite{Planck2018}, and $A_- = 10^{-3}A_+$, such that the power spectrum peaks at $\mathcal{P}_{\mathcal{R}} \sim 5 \times 10^{-3}$.
 The slow-roll parameters~\eqref{eq:hubble_flow_parameters} and the power spectrum~\eqref{eq:power_spectrum} are shown in \fig{fig:starobinsky_set_up}. The range of modes that left the Hubble radius after the transition, and whose homogeneous behaviour we can identify at Hubble exit using our semi-analytical approach~\eqref{eq:covariance_constant_nu_bessel_matched}, is displayed with the black dashed lines. As $\nu$ is approximately constant for the entire post-transition phase, the criterion in \eq{eq:seperate_universe_consistancy_relation} is always met in that range, see \App{app:noise_model_accuracy}.

\begin{figure}
\begin{center}
        \includegraphics[width=0.46\textwidth]{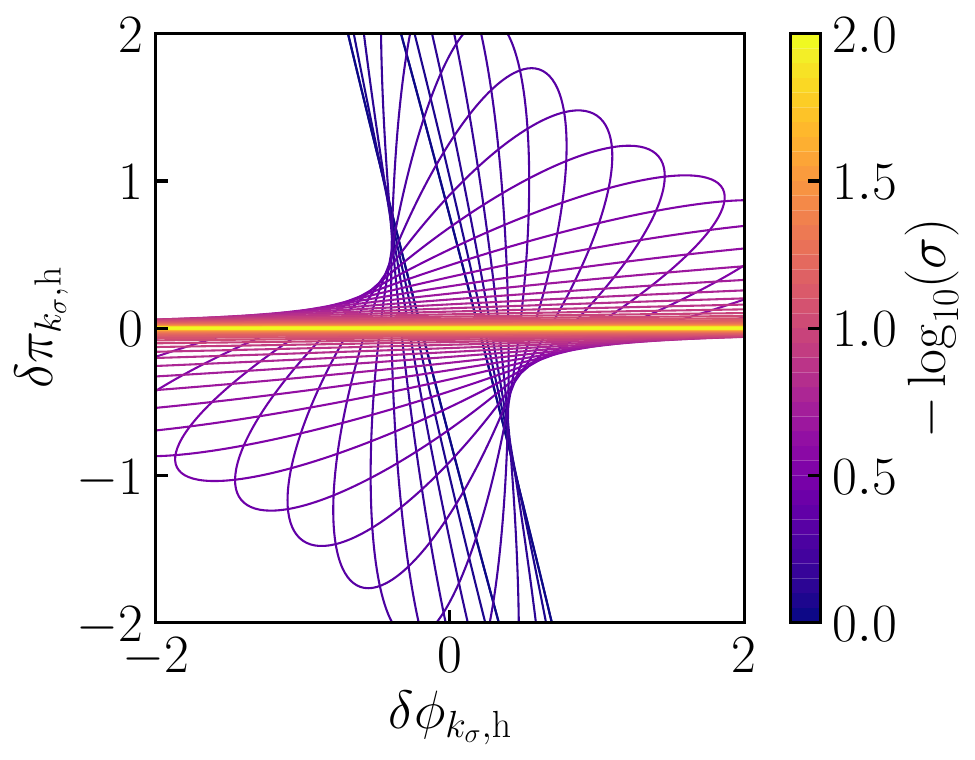}
        \includegraphics[width=0.53\textwidth]{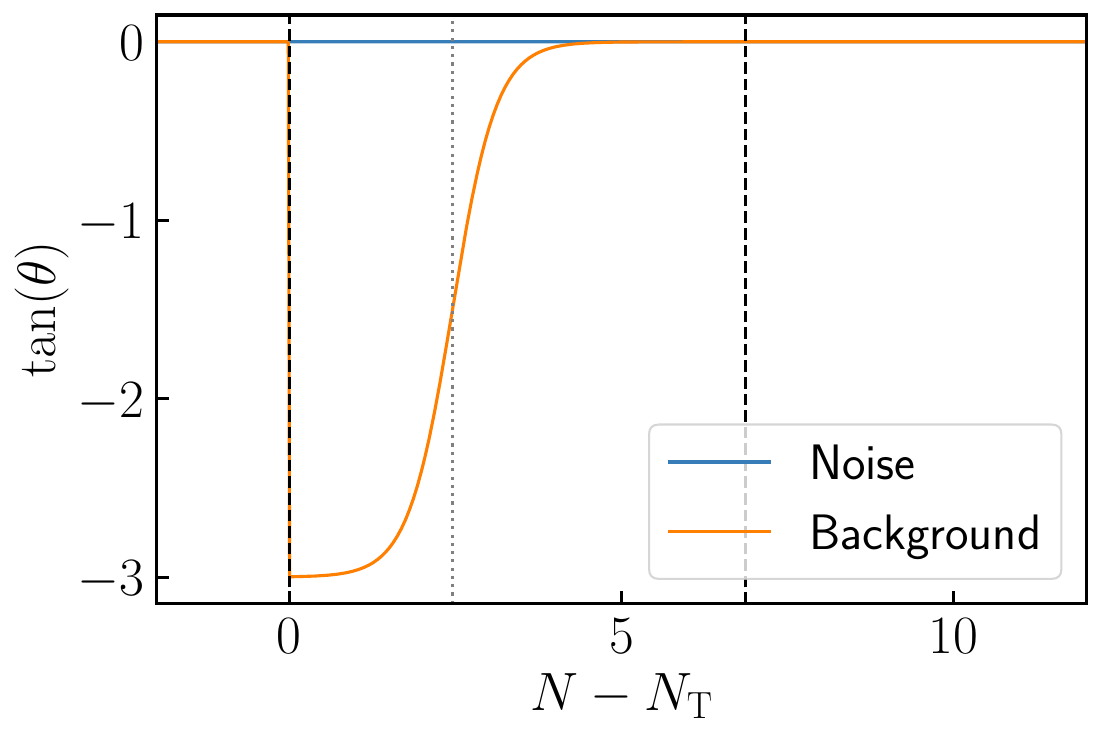}
        \caption{Left: evolution of the ellipse associated with the covariance matrix~\eqref{eq:covariance_matrix} with coarse-graining scale $\sigma$, for the piece-wise linear potential of Sec.~\ref{sub:starobinsky}. The mode shown exited the Hubble radius immediately after the transition with $k= 1.1 \, k_{\mathrm{T}}$. The areas of each ellipse is normalised to $\pi$ to allow visual comparison. Right: the tangent of the phase-space angle for the background, $\partial_N^2 \phi/ \partial_N \phi$, and for the 1D noise found using \eq{eq:covariance_constant_nu_bessel_matched} near the sudden transition at $N_{\mathrm{T}}$. The dotted gray vertical line corresponds to the mode that left the Hubble radius when ultra-slow roll ended and the dashed black lines bound the modes used in the stochastic simulations.}
        \label{fig:starobinsky_squeezing}
\end{center}
\end{figure}

In the left-hand plot of \fig{fig:starobinsky_squeezing}, the ellipses corresponding to the covariance matrix~\eqref{eq:covariance_matrix}
are shown at different coarse-graining scales. The scale shown leaves the Hubble scale immediately after the sudden transition and is soon dominated by the growing mode. This is why it enters a highly-squeezed state, as the eccentricity of the ellipses tends to one. This squeezing occurs even quicker for subsequent modes leaving the Hubble radius. Although this provides an illustrative justification for our use of a 1D treatment of the noise after the transition, the squeezing amplitude (\ie the eccentricity of the ellipse) is not a symplectic invariant quantity and it can be changed by a mere field rescaling, or in general by any canonical transformation~\cite{Grain:2017dqa}. This is why, in practice, we use the criterion given in \eq{eq:seperate_universe_consistancy_relation} to determine when the growing mode dominates and the noise is effectively 1D. 

In the right-hand plot of \fig{fig:starobinsky_squeezing}, the tangent of the phase-space angle of both the background dynamics, $\partial_N \pi/ \partial_N \phi = \epsilon_2/2$, and the 1D noise~\eqref{eq:covariance_constant_nu_bessel_matched} are shown. While at late times the two agree and the phase space is truly 1D, during ultra-slow roll there is a misalignment between the noise and the background, justifying the need to simulate the full 2D phase space.

\begin{figure}
\begin{center}
        \includegraphics[width=\halffigurewidth\textwidth]{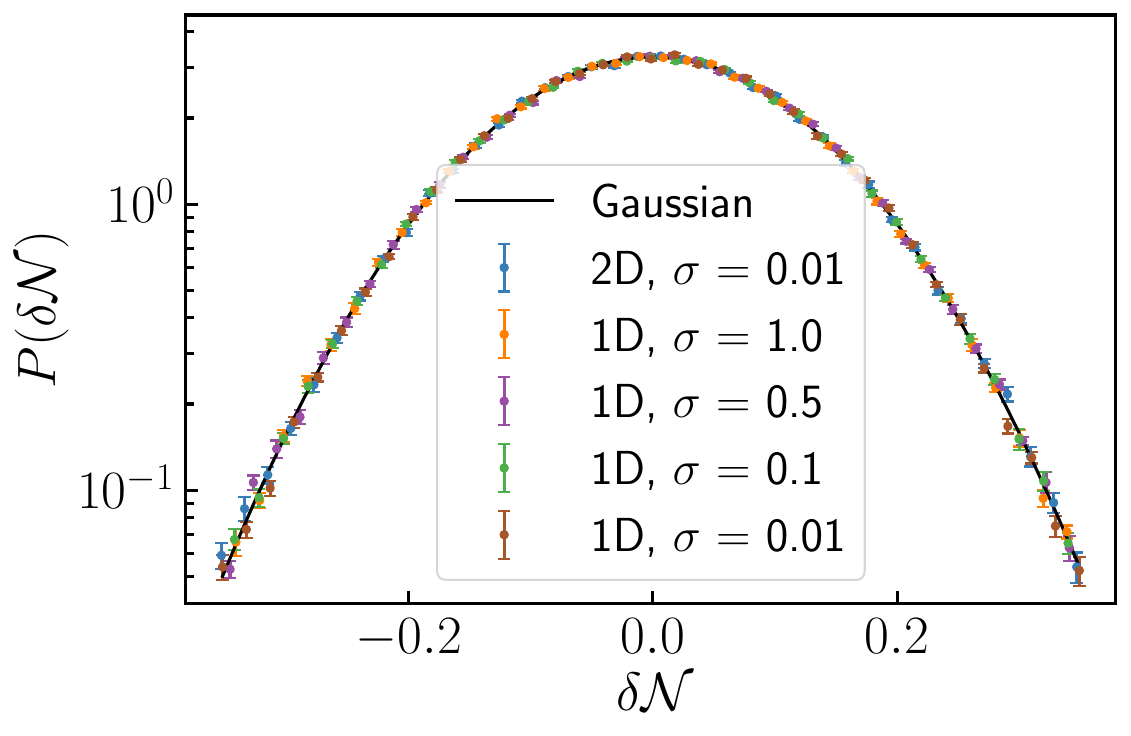}
        \includegraphics[width=\halffigurewidth\textwidth]{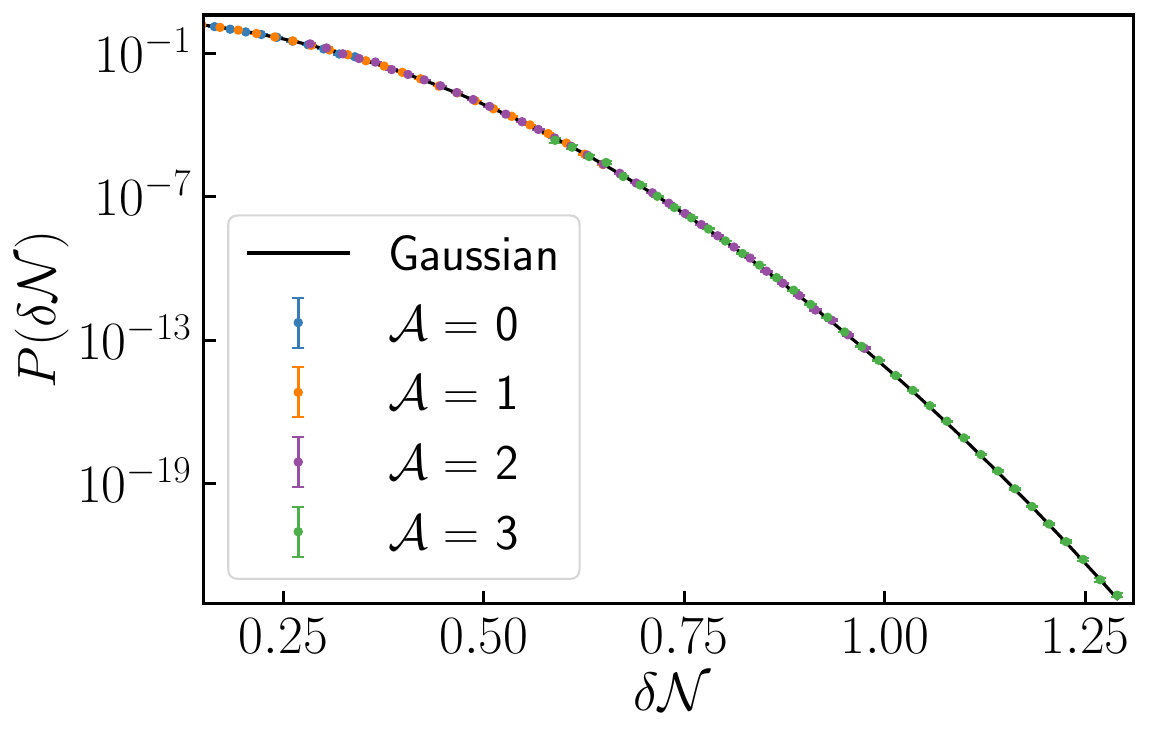}
        \caption{The numerically reconstructed first-passage-time distribution for the piece-wise linear potential~\eqref{eq:potential_starobinsky}.
        The semi-analytical noise model described in \Sec{sub:semi_analytical} is used. Left: comparison of the 1D~\eqref{eq:stochastic_eom_1D} and 2D~\eqref{eq:full_langevin_equation} models. The value of $\sigma = k/(aH)$ defines when the noise is applied and the start/end field values. The solid black curve is Gaussian with variance given by \eq{eq:variance_zeta_lin}. Right: importance sampling using the 1D noise approach~\eqref{eq:importance_sampling_1D_noise_step} to find the tail of the distribution for a range of bias amplitudes, given by the different values of $\mathcal{A}$ shown, for $\sigma=1$. The numerical error bars correspond to one standard deviation, as discussed in Sec.~\ref{sub:numerical_approach}.}
        \label{fig:starobinsky_peak_independance}
\end{center}
\end{figure}

In \fig{fig:starobinsky_peak_independance} we investigate the semi-analytical noise model of \Sec{sub:semi_analytical} combined with the numerical-simulation strategy of \Sec{sub:numerical_approach}. Around the peak of the first-passage time distribution, see the left-hand plot, we run direct sampling stochastic simulations, either using the 1D noise~\eqref{eq:stochastic_eom_1D} or the full 2D noise~\eqref{eq:full_langevin_equation} approaches, for a range of $\sigma$ values. As $\sigma$ is varied, so are $\phi_{\mathrm{init}}/\pi_{\mathrm{init}}$ and $\phi_{\mathrm{end}}$, to correspond to when the first and last mode in the range shown by the vertical dashed lines in \fig{fig:starobinsky_set_up} left the coarse-graining scale. The result is found to be mostly independent of $\sigma$, even for values up to $\sigma=1$, and to agree with the expectation from linear perturbation theory displayed with the black solid curve. This suggests that for modes that left the Hubble radius post-transition, the resulting noise can be modelled using the growing mode at Hubble-crossing, and this validates the noise modelling proposed in \Sec{sub:semi_analytical}. This is also consistent with the fact that the 1D treatment of the noise gives an excellent approximation of the full 2D treatment, see the blue and brown data, in agreement with the discussion around the left panel of \fig{fig:starobinsky_squeezing}, as well as previous works which used $\sigma=0.01$~\cite{Tomberg:2022mkt, Mishra:2023lhe}.

The right-hand plot of \fig{fig:starobinsky_peak_independance} shows the results of applying importance sampling using the 1D noise approach~\eqref{eq:importance_sampling_1D_noise_step} at Hubble crossing, with $\sigma=1$. As the bias amplitude is increased and the numerics probe further into the tail, each data set overlaps with the previous one, including direct sampling with $\mathcal{A}=0$, showing that importance sampling is self-consistent. This confirms our ability to probe the tail of the distribution with importance sampling.

Let us also notice that the Gaussian distribution obtained from standard perturbation theory, whose variance is given in \eq{eq:variance_zeta_lin}, provides a reliable fit to the whole distribution, both around the maximum and on the far tail. This is because, sufficiently long after the transition, there is a linear relationship between $N$ and $\phi$, and thus between $\delta \N$ and $\delta \phi$. Therefore the classical $\delta \Nb$ formalism~\eqref{eq:delta_N_pdf_delta_phi} gives a Gaussian distribution, and we find the stochastic simulations not to deviate substantially from that expectation. Let us however stress that, since a fixed range of comoving modes is integrated over, the non-linear curvature perturbation $\delta\Nb$ is coarse-grained at a scale that exits the Hubble radius in the late slow-roll phase, where the linear relationship between $N$ and $\phi$ holds. 

\subsection{Gaussian bump}
\label{sub:swagat}

\begin{figure}
\begin{center}
        \includegraphics[width=0.48\textwidth]{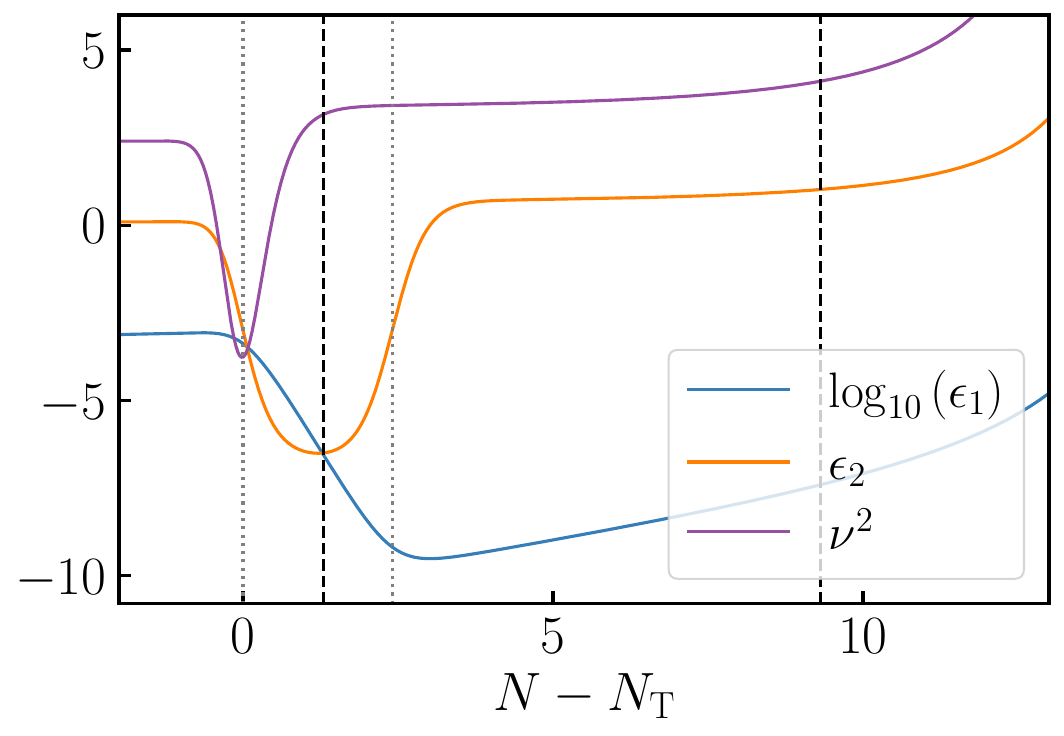}
        \includegraphics[width=0.51\textwidth]{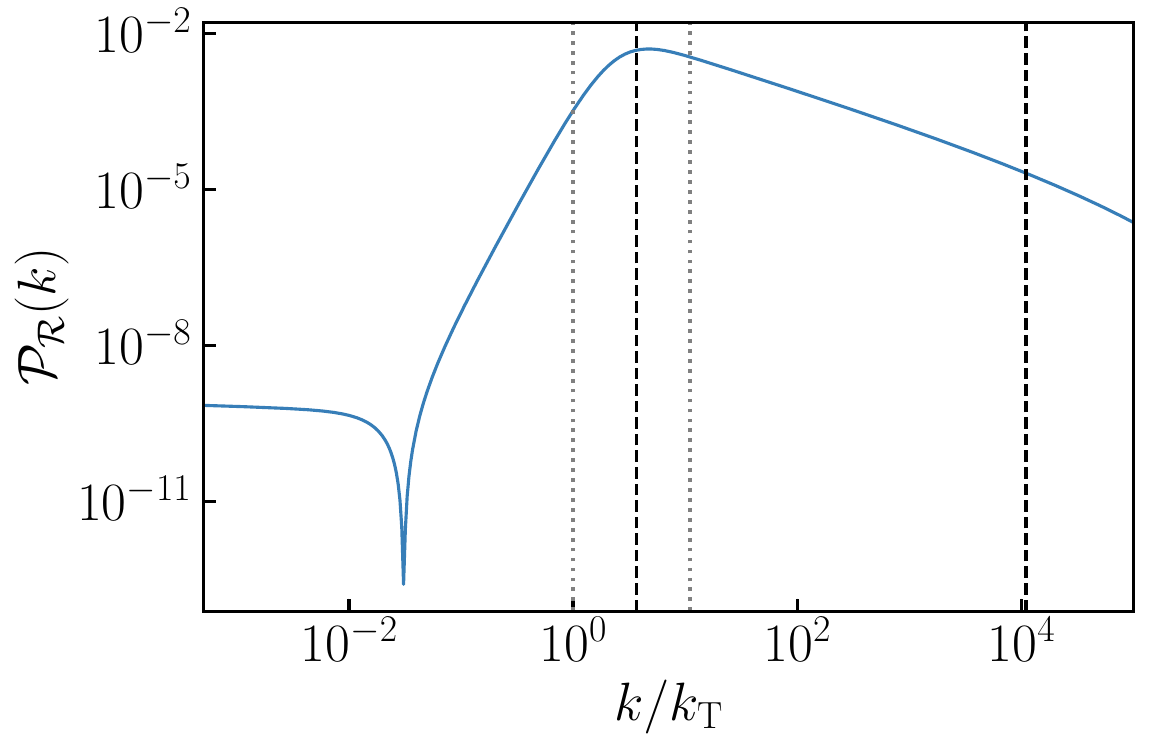}
        \caption{Left: the Hubble-flow parameters~\eqref{eq:hubble_flow_parameters} and $\nu^2$ defined in \eq{eq:nu_squared} for the Gaussian-bump potential~\eqref{eq:potential_swagat}, near the sudden transition at $N_{\mathrm{T}}$, corresponding to the start of ultra-slow roll with $\epsilon_2<-3$. Right: the associated power spectrum~\eqref{eq:power_spectrum} where $k_{\mathrm{T}}$ is the scale that crosses the Hubble radius at time $N_{\mathrm{T}}$. The dotted gray vertical lines frame the modes that left the Hubble radius during ultra-slow roll $(\epsilon_2<-3)$ and the dashed black lines bound the modes integrated over for the noise, as in \fig{fig:starobinsky_set_up}.}
        \label{fig:swagat_set_up}
\end{center}
\end{figure}

Now let us consider a more realistic model with a smooth transition from slow roll to ultra-slow roll. One such model is made of a slow-roll potential with the addition of a Gaussian bump~\cite{Mishra:2019pzq, Mishra:2023lhe},
\begin{equation}
\label{eq:potential_swagat}
    V(\phi) = V_0 \frac{\phi^2}{m^2 + \phi^2}\left\lbrace1 + K\exp \left[ -\frac{1}{2}\frac{(\phi - \phi_0)^2}{\Sigma^2} \right] \right\rbrace \, ,
\end{equation}
where we use $K=1.17 \times 10^{-3}$, $m = 0.5$, $\Sigma = 1.59 \times 10^{-2}$ and $\phi_{0}=2.18812$. These parameters are chosen such that CMB constraints~\cite{Planck2018} are obeyed and the power spectrum peaks at $\mathcal{P}_{\mathcal{R}} \sim 5 \times 10^{-3}$ within the asteroid mass window~\cite{Cole:2023wyx}. In \fig{fig:swagat_set_up} the background dynamics and power spectrum are shown, along with the range of $k$ modes integrated over to compute the noise, $k_{\mathrm{min}}$ and $k_{\mathrm{max}}$. This range is chosen such that the linear power spectrum consistency relation~\eqref{eq:seperate_universe_consistancy_relation} is obeyed for $\sigma \in [0.01, 1]$ for the semi-analytical noise model given in \eq{eq:covariance_constant_nu_bessel_matched}, see Appendix~\ref{app:noise_model_accuracy}. While excluding half of the modes from the ultra-slow-roll phase, this range still includes >85\% of the contribution of the linear power spectrum to $\sigma_{\R}$ when calculated over the full $k$ range shown.

\begin{figure}
\begin{center}
        \includegraphics[width=\halffigurewidth\textwidth]{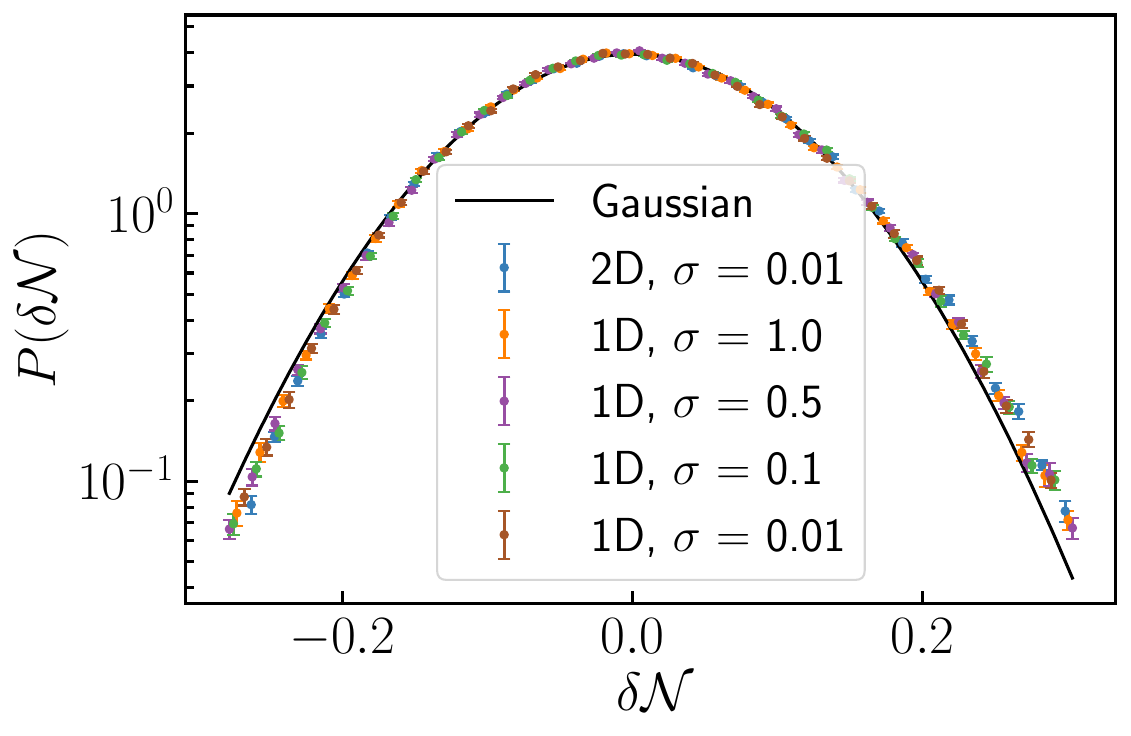}
        \includegraphics[width=\halffigurewidth\textwidth]{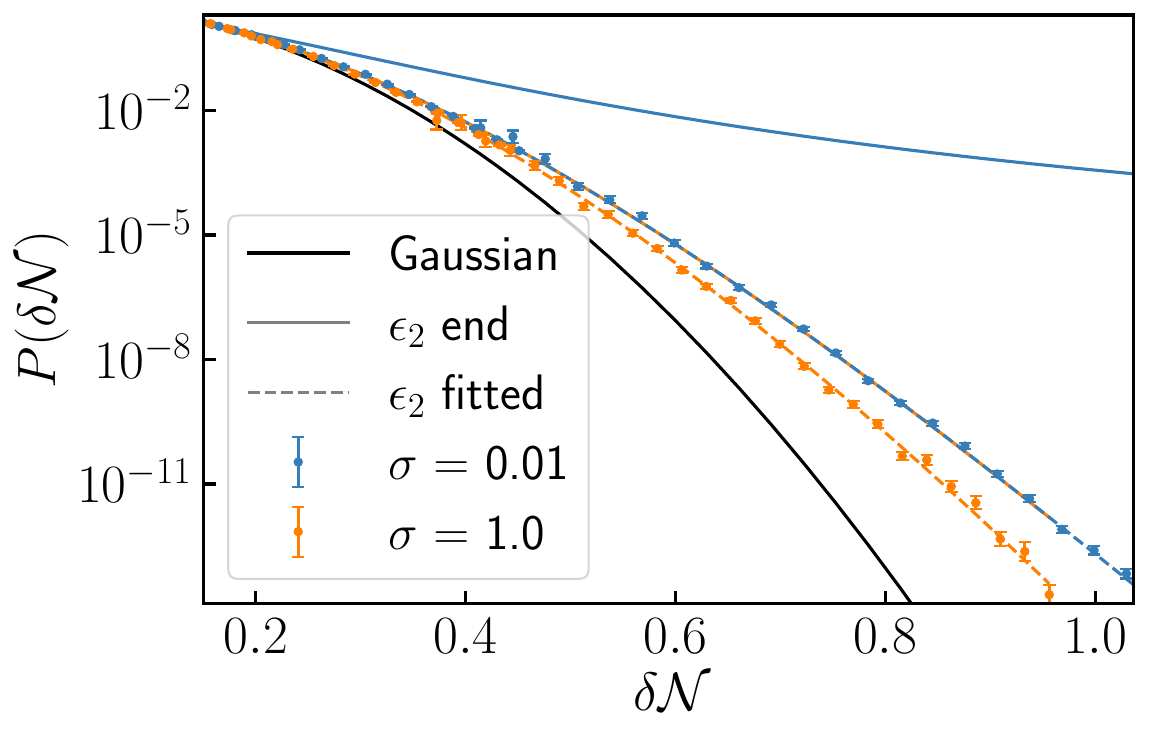}
        \caption{The numerically reconstructed first-passage time distribution for the Gaussian bump potential~\eqref{eq:potential_swagat}. The semi-analytical noise model described in Sec.~\ref{sub:semi_analytical} is used. Left: comparison of the 1D~\eqref{eq:stochastic_eom_1D} and 2D~\eqref{eq:full_langevin_equation} models. The value of $\sigma = k/(aH)$ defines when the noise is applied and the start/end field values. The solid black curve is Gaussian with variance given by \eq{eq:variance_zeta_lin}, \ie it is the result of standard perturbation theory. Right: importance sampling using the 1D noise approach~\eqref{eq:importance_sampling_1D_noise_step} to find the tail of the distribution for a range of bias amplitudes combined, for two $\sigma$ values shown, along with the classical $\delta \Nb$ prediction assuming constant $\epsilon_2$~\eqref{eq:delta_N_pdf_constant_epsilon_2}, colour coded to match. The solid lines used the $\epsilon_2$ value at coarse-graining crossing for the end of the integrated mode range, and the dashed lines are fitted values. The numerical error bars correspond to one standard deviation, as discussed in Sec.~\ref{sub:numerical_approach}. By coincidence, the $\sigma=1$ solid curve and the $\sigma=0.01$ dashed curve overlap.}
\label{fig:swagat_constant_nu_approx_numerics}
\end{center}
\end{figure}

In the left-hand plot of \fig{fig:swagat_constant_nu_approx_numerics} we investigate the semi-analytical noise model of Sec.~\ref{sub:semi_analytical} combined with the numerical strategy of \Sec{sub:numerical_approach}. The variance of the distribution is insensitive to the choice of the $k_{\mathrm{max}}$ used when it was sufficiently far away from the power spectrum peak mode. The peak of $P(\delta \N)$ and the small near-tail non-Gaussianity are independent (within the error bars) both of the choice of $\sigma$ and whether the 2D or 1D noise model is used. Therefore the 1D importance sampling approach~\eqref{eq:importance_sampling_1D_noise_step} is justified.

In the right-hand plot of \fig{fig:swagat_constant_nu_approx_numerics} the far tail of $P(\delta \N)$ is investigated for two different $\sigma$ choices using 1D importance sampling~\eqref{eq:importance_sampling_1D_noise_step}. For large $\delta \Nb\gtrsim 0.1$ there is a clear deviation from the Gaussian distribution with variance given by linear theory \eq{eq:variance_zeta_lin}. While the peaks of the distributions agree, the exponential-like far tails deviate. At $\delta \N =1$, the $\sigma=1$ and $\sigma =0.01$ results for $P(\delta \N )$ are enhanced by factors of $\sim 10^{6}$ and $\sim 10^{8}$ respectively, compared to the Gaussian prediction.

The classical $\delta \Nb$ prediction~\eqref{eq:delta_N_pdf_constant_epsilon_2} is in agreement with the results of \fig{fig:swagat_constant_nu_approx_numerics} if we choose the value of $\epsilon_2$ appropriately. As $\epsilon_2$ is not exactly constant during this period, some post-hoc fitting is required (the dashed lines). The value of $\epsilon_2$ needed for the $\sigma_1=1$ and $\sigma_2=0.01$ data are $\epsilon_2 = 0.732$ and $\epsilon_2 = 1.02$, respectively. The \efolding times when these values occurred were $N_1-N_{\mathrm{T}} = 4.7$ and $N_2-N_{\mathrm{T}} = 9.3$, respectively, and  were separated by $N_1-N_2=\ln(\sigma_1/\sigma_2)$. Both values occurred during the constant $\epsilon_2$ phase, suggesting an ``effective'' $\epsilon_2$ value corresponding to a time shortly after the power spectrum peak scale crossed the coarse-graining scale. If the value of $\epsilon_2$ corresponding to the coarse-graining crossing time of the last mode integrated is used (solid lines), then the classical $\delta \Nb$ does not fit the data. This is expected for the solid $\sigma=0.01$ curve, as $\epsilon_2$ is not constant when the coarse-graining crossing of that scale occurred. 

\begin{figure}
\begin{center}
        \includegraphics[width=\halffigurewidth\textwidth]{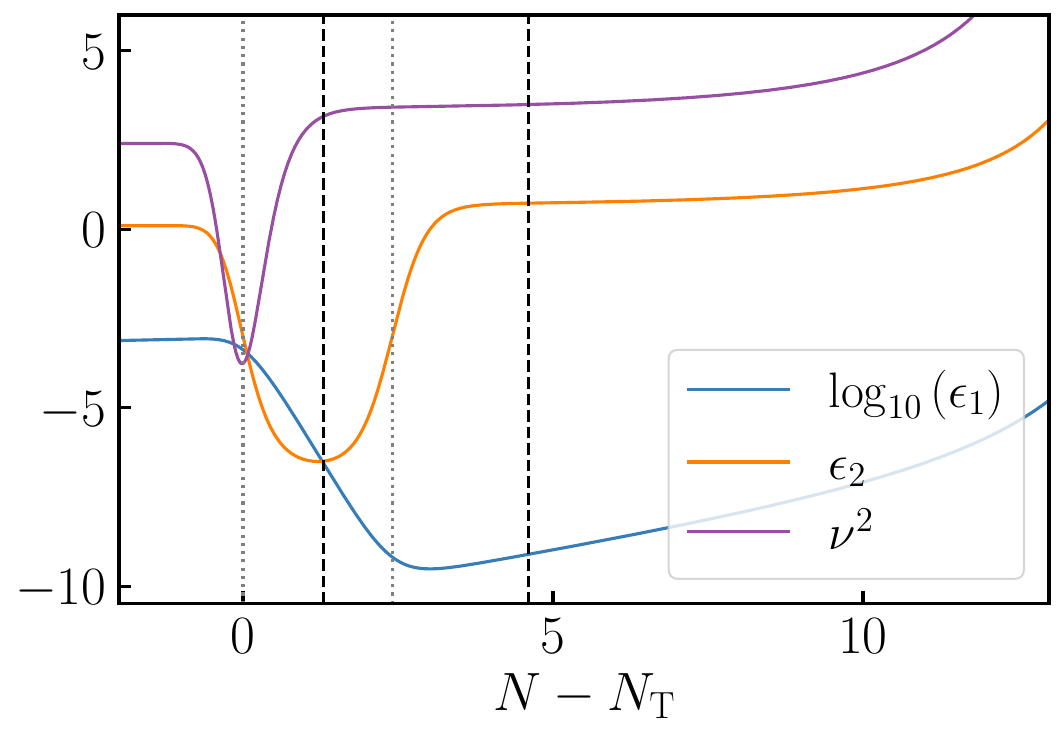}
        \includegraphics[width=\halffigurewidth\textwidth]{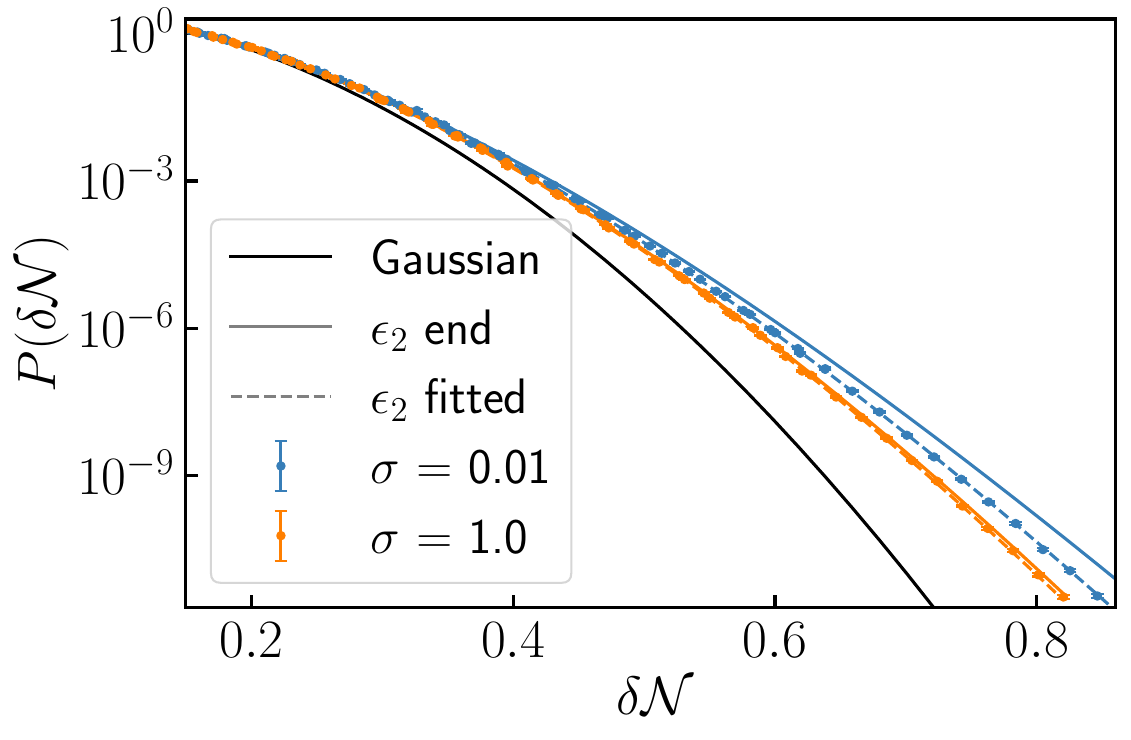}
        \caption{Left: the same as the left-hand plot of \fig{fig:swagat_set_up} but $k_{\mathrm{max}}$ has been shifted to the constant $\epsilon_2$ phase. Right: the same as the right-hand plot of \fig{fig:swagat_constant_nu_approx_numerics} but for the mode range shown on the left panel.}
        \label{fig:swagat_constant_nu_approx_numerics2}
\end{center}
\end{figure}

If however the range of modes integrated, depicted in \fig{fig:swagat_set_up}, is changed such that the last mode integrated left the coarse-graining scale during the constant-$\epsilon_2$ phase, then the classical $\delta \Nb$ prediction~\eqref{eq:delta_N_pdf_constant_epsilon_2} provides a reasonable fit to the data. This is shown in \fig{fig:swagat_constant_nu_approx_numerics2} for $\sigma=1$ data, with both classical $\delta \Nb$ curves giving a reasonable fit. The $\sigma=0.01$ data still requires a fitted $\epsilon_2$ but the classical $\delta N$ curve with $\epsilon_2$ value corresponding to when the last mode left the coarse-graining scale is a much closer fit than for \fig{fig:swagat_constant_nu_approx_numerics}.

We see an increased scatter in the importance sampling data in \fig{fig:swagat_constant_nu_approx_numerics} compared to \figs{fig:starobinsky_peak_independance} and \ref{fig:swagat_constant_nu_approx_numerics2}. This is because the correlation between the weights, $\ln{\left(w\right)}$, and the value of $\delta \mathcal{N}$ is weaker, increasing the error when the lognormal estimator is used. See \Refa{Jackson:2022unc} for details.

\section{Conclusions}
\label{sec:conclusion}

In this paper we have, for the first time, shown how to numerically investigate the far tail of the first-passage-time probability distribution, $P(\delta \N)$, using importance sampling in inflation models including a transient period of ultra-slow roll following a sudden transition. We investigated two specific inflaton potentials leading to strongly enhanced power spectra on small scales, $\mathcal{P}_{\mathcal{R}} \sim 5 \times 10^{-3}$, as typically studied in single-field models of primordial black hole production from inflation~\cite{LISACosmologyWorkingGroup:2023njw}. To do so we have extended the open-source \pyfpt code to describe the stochastic dynamics in a 2D phase space, \eq{eq:stochstic_eom}, incorporating noise that departs from the standard Bunch--Davies profile due to non-trivial sub-Hubble dynamics during the transition. We have applied importance sampling techniques and a 1D noise model, see \eq{eq:importance_sampling_1D_noise_step}. The noise was pre-computed on the classical background for efficiency. With this approach, we have been able to explore the probability distribution down to $P(\delta \N)\lesssim 10^{-10}$ in simulations taking less than 15 minutes to run on a laptop computer.

We have demonstrated that a 1D model for the noise was justified by the squeezing in phase space of the field perturbations on super-Hubble scales even after a sudden transition to ultra-slow roll, as was shown in \fig{fig:starobinsky_squeezing}. In our simulations we required that keeping only the growing mode of the homogeneous ($k=0$) field perturbation found at coarse graining, \eq{eq:separate_universe_power_spectrum}, reproduced within 1\% the linear power spectrum at the end of inflation obtained by evolving the full inhomogeneous field perturbation, \eq{eq:power_spectrum}, see Appendix~\ref{app:noise_model_accuracy}. This ensured the 1D noise model was consistent with linear perturbation theory for all the modes used in the simulations and the separate-universe approach was correctly applied~\cite{Jackson:2023obv}. 

In general it is non-trivial to identify the long-wavelength limit of the field perturbation from the numerical solution of the full mode equation at a finite wavelength at the coarse-graining time. One possibility is simply to wait long enough (long after Hubble exit) until the wavelength is large enough that the spatial gradients are negligible. However we showed that it was also possible to accurately identify the long-wavelength limit by matching the full numerical solution to a Bessel function ansatz soon after, or even at, Hubble exit. We were then able to faithfully reproduce the linear power spectrum even when coarse graining at the Hubble scale, $k=aH$. For a piece-wise linear potential, see Sec.~\ref{sub:starobinsky}, our stochastic numerical results matched the classical Gaussian prediction far into the tail of the PDF (which we expect to be an excellent approximation in this linear case), showing the validity of the 1D noise approach.

In Sec.~\ref{sub:swagat} we then applied our \pyfpt numerical code to a smooth inflaton potential with a small Gaussian bump, leading to a sudden transition to ultra-slow roll, on top of a plateau-type potential, compatible with CMB constraints on large-scale perturbations~\cite{Mishra:2019pzq}. The probability distribution, $P(\delta \N)$, is well-described by a Gaussian close to the peak of the distribution, but we found strong deviations from Gaussianity and an exponential tail for $P(\delta \N)\lesssim 10^{-5}$. A simple analytic function~\eqref{eq:delta_N_pdf_constant_epsilon_2} gave an excellent fit to the distribution. This form for the PDF is consistent with the classical $\delta\Nb$ formalism~\cite{Starobinsky:1986fx,Sasaki1996,Sasaki:1998ug,Wands:2000dp,Lyth:2004gb} in constant-roll inflation (where $\epsilon_2=$ constant), in agreement with previous studies~\cite{Tomberg:2023kli, Ballesteros:2024pwn}. Since $\epsilon_2$ is not exactly constant in this model, there is some ambiguity in the precise value of $\epsilon_2$ used in \eq{eq:delta_N_pdf_constant_epsilon_2} introduced by the time variation, which is seen in \fig{fig:swagat_constant_nu_approx_numerics}. If however we stop the simulation when the variation of $\epsilon_2$ is very small, then the classical $\delta \Nb$ provides an excellent fit, as seen in \fig{fig:swagat_constant_nu_approx_numerics2}. 

Let us also stress that in the Gaussian-bump model, the tail behaviour was found to depend quite significantly on the value of $\sigma$, \ie on the ratio between the coarse-graining and the Hubble scale. In principle, $\sigma$ should be set such that the coarse-graining scale is the one at which backreaction effectively takes place in the local FLRW patches. There is no clear prescription for where that should be, and this is somehow inherent to the accuracy of the separate-universe picture itself, but values of order one for $\sigma$ seem reasonable (\ie assuming that backreaction occurs at the Hubble scale). However, to correctly identify the long-wavelength limit and extract the asymptotic growing-mode may require us to wait long enough after Hubble exit such that the spatial gradients are negligible, and this is why values of $\sigma\ll 1$ are usually employed in stochastic inflation. This is indeed required in standard modelling of the noise, but in regimes where $\nu$ is approximately constant, the semi-analytical approach we have proposed was found to work accurately even at values of $\sigma = 1$. This makes it particularly well-suited to cases where a substantial dependence on $\sigma$ is found, as seems to be the case for non-Gaussian tails, where values of $\sigma\ll 1$ might underestimate backreaction effects.

The focus of this work was to propose a noise modelling and to design an importance-sampling approach that allows one to study post-transition phases during inflation, where slow roll is violated. This is why we focused on the statistics of the first-passage time arising from a fixed range of comoving scales, and computed the noise along a fixed background reference trajectory, to simplify the discussion and prevent more subtle stochastic effects to interfere in the analysis. This led us to situations where the stochastic-$\delta \Nb$ results seem to be easily interpreted in terms of classical-$\delta \Nb$ effects only. Although some works suggest that the heavy tails found in the stochastic-$\delta \Nb$ formalism come from purely classical effects, with results from classical- and stochastic-$\delta\Nb$ formalisms in excellent agreement~\cite{Ballesteros:2024pwn}, let us stress that this cannot be true in general. The example considered in \Sec{sub:starobinsky} is a good counter-example: since the classical dynamics gives rise to a linear relationship between $\phi$ and $N$ at late time, a Gaussian distribution is necessarily found. However, the full stochastic-$\delta \Nb$ formalism always yields exponential tails~\cite{Pattison:2017mbe, Ezquiaga:2019ftu}, due to each realisation integrating over a different range of modes (here we integrated over the same range), so the two approaches necessarily deviate. In addition, the curvature perturbation is never measured in terms of comoving scales, but rather in terms of physical scales. In the classical approach, going from comoving to physical scales is straightforward (although it can lead to non-trivial effects in more than one-dimensional phase spaces~\cite{Tada:2016pmk}) since there is a one-to-one correspondence between scales and field configurations when they exit the Hubble radius. In the stochastic picture however, this is no longer the case, convolutions across the whole inflating domain need to be performed~\cite{Tada:2021zzj, Ando:2020fjm, Animali:2024jiz}. This leads to drastic changes in observables involving large fluctuations that cannot be accounted for by the classical approach.

To include non-Gaussianties in the tail of $P(\delta \N)$ due to stochastic effects beyond the classical $\delta \Nb$ approximation, one would need to go beyond pre-computing the noise using the classical background trajectory, \ie one should include backreaction of the stochastic evolution on the noise itself. In principle this would involve solving the mode equation for quantum fluctuations at every time-step in every realisation which would be computationally prohibitive and lead to non-Markovian noise. However we have seen that in the decoupling limit (when $\epsilon_1\to0$) during ultra-slow roll, the noise at coarse graining could be described as a function of the local field value, using \eq{eq:covariance_constant_nu_bessel_matched} and \eq{eq:nu_squared_approximation}, independent of the particular classical phase-space trajectory, see \fig{fig:swagat_phase_space}. Hence the local field value at each step could be used to estimate the noise including the stochastic backreaction in each realisation. This has the advantage of not requiring the mode equations to be solved on each stochastic trajectory separately, greatly speeding up the numerics. It is still an open question if this approach is valid for a fully stochastic background, and we leave the investigation of this for future work. Additionally, how best to find the potentially $k$-dependent amplitude of the growing mode in this approach (the Bessel coefficient $B_k$ in \eq{eq:covariance_constant_nu_bessel_matched}) requires further consideration, which we also leave for future work.

Due to the sudden transition from slow roll to ultra-slow roll leading to a discontinuity in the behaviour of the homogeneous part of super-Hubble modes~\cite{Jackson:2023obv}, we have only simulated the stochastic evolution after the transition. It remains an unsolved challenge to build a full stochastic model of super-Hubble modes through a sudden transition (\ie in phases where $\nu$ varies quickly). Nonetheless in this work we have demonstrated the persistence of exponential tails in the probability distribution $P(\delta \N)$ incorporating non-slow-roll dynamics and non-adiabatic perturbations immediately after a sudden transition, including the peak of the primordial power spectrum and more than 85\% of the total power integrated over all observable scales. 

Finally we note that while the diffusion-based bias model~\eqref{eq:2D_1D_noise_diussion_based_bias} used for importance sampling was able to probe the far tail of $P(\delta \Nb)$ for the potentials investigated, it might not be sufficient in more general cases. This is due to the quality of the importance sampling data (the correlation between $\ln (w)$ and $\delta \Nb$, see Ref~\cite{Jackson:2022unc}) depending on the choice of bias and the model investigated. The problem of finding the optimal bias, which gives the highest quality data, has been solved in 1D~\cite{Tomberg:2022mkt} but is still an important open question in the 2D phase space. 
\section*{Data Availability Statement} 

The data from which the results presented here are derived is freely available at \url{https://github.com/Jacks0nJ/PyFPT}, along with the code required to produce the data.

\acknowledgments

The authors would like to thank Parth Bhargava, Laura Iacconi, Swagat Mishra, Shi Pi and Eemeli Tomberg for insightful discussions. This work was supported by the Science and Technology Facilities Council (grant numbers ST/T506345/1 and ST/W001225/1). For the purpose of open access, the authors have applied a Creative Commons Attribution (CC-BY) licence to any Author Accepted Manuscript version arising from this work.

\appendix
\section{Noise model accuracy}
\label{app:noise_model_accuracy}

\begin{figure}[ht]
\begin{center}
        \includegraphics[width=\halffigurewidth\textwidth]{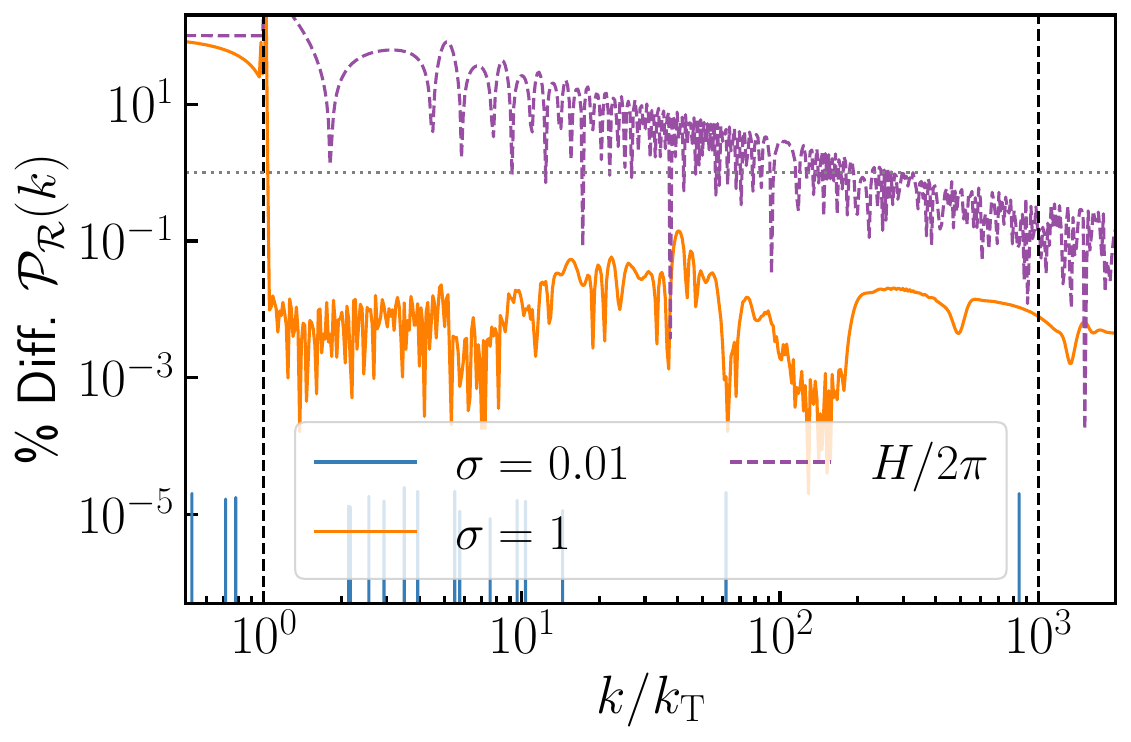}
        \includegraphics[width=\halffigurewidth\textwidth]{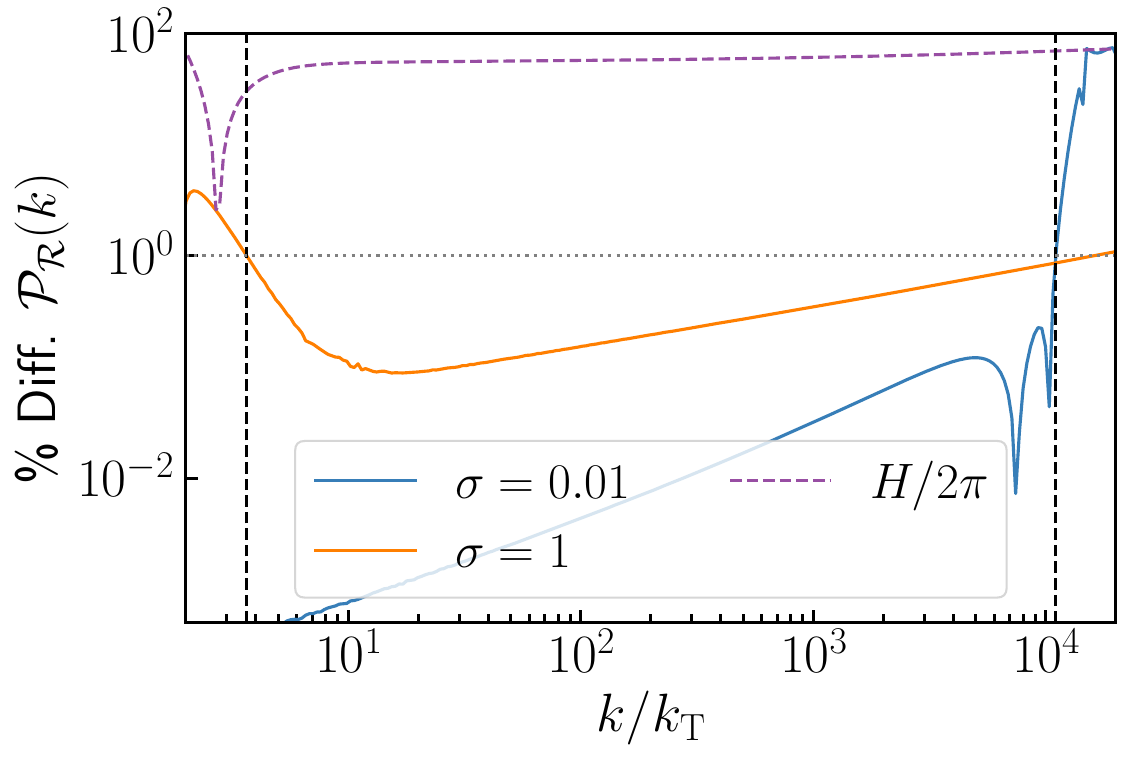}
        \caption{The percentage difference in the linear power spectrum at the end of inflation, $\mathcal{P}_{\mathcal{R}}(k)$, obtained using different models for the noise 
        to determine the amplitude of the homogeneous growing mode, $C_k$ in \eq{eq:separate_universe_power_spectrum}, against the true value found by solving the full mode equation~\eqref{eq:full_delta_phi} for $\delta\phi_k$ in \eq{eq:power_spectrum} at the end of inflation. The solid curves correspond to the semi-analytical procedure, detailed in Sec.~\ref{sub:semi_analytical}, to identify the homogeneous part of the field fluctuations when $k=\sigma aH$, while the dashed curve corresponds to the simple analytical noise for the Bunch--Davies vacuum state for a massless field in de Sitter~\eqref{eq:covariance_de_sitter} leading to typical field fluctuations $H/2\pi$ at Hubble exit.  The homogeneous growing mode is then calculated at the matching time using \eq{eq:separate_universe_constants}. The vertical dashed black lines indicate the range of modes used in our stochastic simulations, corresponding to $k_{\mathrm{min}}$ on the left and $k_{\mathrm{max}}$ on the right. The horizontal dotted gray line is the 1\% accuracy threshold. Left: the piece-wise linear potential~\eqref{eq:potential_starobinsky}. Right: the Gaussian-bump model~\eqref{eq:potential_swagat}.}
        \label{fig:constant_nu_approx_numerics_percent_diff}
\end{center}
\end{figure}

In this appendix, the requirement that the homogeneous perturbations used to calculate the stochastic noise, $\delta \phi_{k_\sigma,\mathrm{h}}$ and $\delta \pi_{k_\sigma,\mathrm{h}}$ in \eq{eq:noise_moments_simplified}, give the correct linear power spectrum at the end of inflation using the separate-universe approach~\eqref{eq:seperate_universe_consistancy_relation}, is tested for the two potentials investigated. Here the homogeneous perturbations were found using the semi-analytical Bessel-function method of Sec.~\ref{sub:semi_analytical}, and keeping only the homogeneous growing mode, see \eq{eq:covariance_constant_nu_bessel_matched}.

The results are shown for the piece-wise linear~\eqref{eq:potential_starobinsky} and Gaussian bump~\eqref{eq:potential_swagat} potentials in the left and right panels respectively of \fig{fig:constant_nu_approx_numerics_percent_diff}. For the piece-wise linear potential, the criterion given in \eq{eq:seperate_universe_consistancy_relation} is always met after the sudden transition. Therefore a range of modes to include the majority of the oscillations in the power spectrum post-transition, see \fig{fig:starobinsky_set_up}, is chosen for the simulations. For the Gaussian-bump potential, the criterion is met for $0.01\leq \sigma \leq 1$ for the range indicated by the vertical dashed lines. Note that using the standard $H/2\pi$ field fluctuations for a massless field in de Sitter to calculate the noise, \ie using \eq{eq:covariance_de_sitter}, gives a $\gtrsim10\%$ error close to the transition for the piece-wise linear potential and $\sim100\%$ error throughout for the Gaussian bump potential, showing the importance of improving upon this simple analytical approximation.

\bibliographystyle{JHEP-edit}
\bibliography{main.bib}

\providecommand{\href}[2]{#2}\begingroup\raggedright\begin{thebibliography}{10}

\bibitem{Starobinsky:1980te}
A.~A. Starobinsky, \emph{A new type of isotropic cosmological models without
  singularity},
  \href{https://doi.org/10.1016/0370-2693(80)90670-X}{\emph{Physics Letters B}
  {\bfseries 91}{\bfseries (1)} (1980) 99}.

\bibitem{Sato:1980yn}
K.~Sato, \emph{{First-order phase transition of a vacuum and the expansion of
  the Universe}}, \href{https://doi.org/10.1093/mnras/195.3.467}{\emph{Monthly
  Notices of the Royal Astronomical Society} {\bfseries 195}{\bfseries (3)}
  (1981) 467}.

\bibitem{Guth:1980zm}
A.~H. Guth, \emph{{Inflationary universe: A possible solution to the horizon
  and flatness problems}},
  \href{https://doi.org/10.1103/PhysRevD.23.347}{\emph{Physical Review D}
  {\bfseries 23}{\bfseries (2)} (1981) 347}.

\bibitem{Linde:1981mu}
A.~D. Linde, \emph{{A new inflationary universe scenario: A possible solution
  of the horizon, flatness, homogeneity, isotropy and primordial monopole
  problems}}, \href{https://doi.org/10.1016/0370-2693(82)91219-9}{\emph{Physics
  Letters B} {\bfseries 108}{\bfseries (6)} (1982) 389}.

\bibitem{Albrecht:1982wi}
A.~Albrecht and P.~J. Steinhardt, \emph{{Cosmology for Grand Unified Theories
  with Radiatively Induced Symmetry Breaking}},
  \href{https://doi.org/10.1103/PhysRevLett.48.1220}{\emph{Physical Review
  Letters} {\bfseries 48}{\bfseries (17)} (1982) 1220}.

\bibitem{Linde:1983gd}
A.~D. Linde, \emph{Chaotic inflation},
  \href{https://doi.org/10.1016/0370-2693(83)90837-7}{\emph{Physics Letters B}
  {\bfseries 129}{\bfseries (3--4)} (1983) 177}.

\bibitem{Planck2018}
{\scshape Planck collaboration}, \emph{{Planck 2018 results. VI. Cosmological
  parameters}},
  \href{https://doi.org/10.1051/0004-6361/201833910}{\emph{Astronomy \&
  Astrophysics} {\bfseries 641} (2020) A6}
  [\href{https://arxiv.org/abs/1807.06209}{{\ttfamily 1807.06209}}].

\bibitem{Tristram:2021tvh}
M.~Tristram et~al., \emph{{Improved limits on the tensor-to-scalar ratio using
  BICEP and Planck data}},
  \href{https://doi.org/10.1103/PhysRevD.105.083524}{\emph{Physical Review D}
  {\bfseries 105}{\bfseries (8)} (2022) 083524}
  [\href{https://arxiv.org/abs/2112.07961}{{\ttfamily 2112.07961}}].

\bibitem{Mukhanov:1981xt}
V.~F. Mukhanov and G.~V. Chibisov, \emph{Quantum fluctuations and a nonsingular
  universe},
  \href{http://www.jetpletters.ru/ps/1510/article_23079.shtml}{\emph{Journal of
  Experimental and Theoretical Physics Letters} {\bfseries 33}{\bfseries (10)}
  (1981) 532}.

\bibitem{Mukhanov:1982nu}
V.~F. Mukhanov and G.~V. Chibisov, \emph{Vacuum energy and large-scale
  structure of the universe},
  \href{http://jetp.ras.ru/cgi-bin/e/index/e/56/2/p258?}{\emph{Journal of
  Experimental and Theoretical Physics} {\bfseries 56}{\bfseries (2)} (1982)
  258}.

\bibitem{Starobinsky:1982ee}
A.~A. Starobinsky, \emph{{Dynamics of Phase Transition in the New Inflationary
  Universe Scenario and Generation of Perturbations}},
  \href{https://doi.org/10.1016/0370-2693(82)90541-X}{\emph{Physics Letters B}
  {\bfseries 117}{\bfseries (3--4)} (1982) 175}.

\bibitem{Guth:1982ec}
A.~H. Guth and S.-Y. Pi, \emph{{Fluctuations in the New Inflationary
  Universe}}, \href{https://doi.org/10.1103/PhysRevLett.49.1110}{\emph{Physical
  Review Letters} {\bfseries 49}{\bfseries (15)} (1982) 1110}.

\bibitem{Hawking:1982cz}
S.~Hawking, \emph{The development of irregularities in a single bubble
  inflationary universe},
  \href{https://doi.org/10.1016/0370-2693(82)90373-2}{\emph{Physics Letters B}
  {\bfseries 115}{\bfseries (4)} (1982) 295}.

\bibitem{Bardeen:1983qw}
J.~M. Bardeen, P.~J. Steinhardt and M.~S. Turner, \emph{Spontaneous creation of
  almost scale-free density perturbations in an inflationary universe},
  \href{https://doi.org/10.1103/PhysRevD.28.679}{\emph{Physical Review D}
  {\bfseries 28}{\bfseries (4)} (1983) 679}.

\bibitem{Chluba:2015bqa}
J.~Chluba, J.~Hamann and S.~P. Patil, \emph{{Features and new physical scales
  in primordial observables: Theory and observation}},
  \href{https://doi.org/10.1142/S0218271815300232}{\emph{International Journal
  of Modern Physics D} {\bfseries 24}{\bfseries (10)} (2015) 1530023}
  [\href{https://arxiv.org/abs/1505.01834}{{\ttfamily 1505.01834}}].

\bibitem{Zeldovich1967}
Y.~B. Zel'dovich and I.~D. Novikov, \emph{{The Hypothesis of Cores Retarded
  during Expansion and the Hot Cosmological Model}}, \emph{Soviet Astronomy}
  {\bfseries 10} (1967) 602.

\bibitem{Hawking:1971ei}
S.~Hawking, \emph{Gravitationally collapsed objects of very low mass},
  \href{https://doi.org/10.1093/mnras/152.1.75}{\emph{Monthly Notices of the
  Royal Astronomical Society} {\bfseries 152} (1971) 75}.

\bibitem{Carr:1974nx}
B.~J. Carr and S.~W. Hawking, \emph{{Black holes in the early Universe}},
  \href{https://doi.org/10.1093/mnras/168.2.399}{\emph{Monthly Notices of the
  Royal Astronomical Society} {\bfseries 168} (1974) 399}.

\bibitem{Carr:2016drx}
B.~Carr, F.~Kühnel and M.~Sandstad, \emph{Primordial black holes as dark
  matter}, \href{https://doi.org/10.1103/PhysRevD.94.083504}{\emph{Physical
  Review D} {\bfseries 94}{\bfseries (8)} (2016) 083504}
  [\href{https://arxiv.org/abs/1607.06077}{{\ttfamily 1607.06077}}].

\bibitem{Carr:2020xqk}
B.~Carr and F.~Kühnel, \emph{{Primordial Black Holes as Dark Matter: Recent
  Developments}},
  \href{https://doi.org/10.1146/annurev-nucl-050520-125911}{\emph{Annual Review
  of Nuclear and Particle Science} {\bfseries 70} (2020) 355}
  [\href{https://arxiv.org/abs/2006.02838}{{\ttfamily 2006.02838}}].

\bibitem{Green:2020jor}
A.~M. Green and B.~J. Kavanagh, \emph{Primordial black holes as a dark matter
  candidate}, \href{https://doi.org/10.1088/1361-6471/abc534}{\emph{Journal of
  Physics G} {\bfseries 48}{\bfseries (4)} (2021) 043001}
  [\href{https://arxiv.org/abs/2007.10722}{{\ttfamily 2007.10722}}].

\bibitem{Green:2024bam}
A.~M. Green, \emph{Primordial black holes as a dark matter candidate - a brief
  overview},
  \href{https://doi.org/10.1016/j.nuclphysb.2024.116494}{\emph{Nuclear Physics
  B} {\bfseries 1003} (2024) 116494}
  [\href{https://arxiv.org/abs/2402.15211}{{\ttfamily 2402.15211}}].

\bibitem{Carr:2019kxo}
B.~Carr, S.~Clesse, J.~García-Bellido and F.~Kühnel, \emph{Cosmic conundra
  explained by thermal history and primordial black holes},
  \href{https://doi.org/10.1016/j.dark.2020.100755}{\emph{Physics of the Dark
  Universe} {\bfseries 31} (2021) 100755}
  [\href{https://arxiv.org/abs/1906.08217}{{\ttfamily 1906.08217}}].

\bibitem{Ananda:2006af}
K.~N. Ananda, C.~Clarkson and D.~Wands, \emph{{The Cosmological gravitational
  wave background from primordial density perturbations}},
  \href{https://doi.org/10.1103/PhysRevD.75.123518}{\emph{Physical Review D}
  {\bfseries 75} (2007) 123518}
  [\href{https://arxiv.org/abs/gr-qc/0612013}{{\ttfamily gr-qc/0612013}}].

\bibitem{Baumann:2007zm}
D.~Baumann, P.~J. Steinhardt, K.~Takahashi and K.~Ichiki, \emph{{Gravitational
  Wave Spectrum Induced by Primordial Scalar Perturbations}},
  \href{https://doi.org/10.1103/PhysRevD.76.084019}{\emph{Physical Review D}
  {\bfseries 76} (2007) 084019}
  [\href{https://arxiv.org/abs/hep-th/0703290}{{\ttfamily hep-th/0703290}}].

\bibitem{EPTA:2023xxk}
{\scshape EPTA, InPTA} collaboration, \emph{{The second data release from the
  European Pulsar Timing Array - IV. Implications for massive black holes, dark
  matter, and the early Universe}},
  \href{https://doi.org/10.1051/0004-6361/202347433}{\emph{Astronomy \&
  Astrophysics} {\bfseries 685} (2024) A94}
  [\href{https://arxiv.org/abs/2306.16227}{{\ttfamily 2306.16227}}].

\bibitem{NANOGrav:2023hvm}
{\scshape NANOGrav collaboration}, \emph{{The NANOGrav 15 yr Data Set: Search
  for Signals from New Physics}},
  \href{https://doi.org/10.3847/2041-8213/acdc91}{\emph{The Astrophysical
  Journal Letters} {\bfseries 951}{\bfseries (1)} (2023) L11}
  [\href{https://arxiv.org/abs/2306.16219}{{\ttfamily 2306.16219}}].

\bibitem{LISACosmologyWorkingGroup:2023njw}
{\scshape LISA Cosmology Working Group} collaboration, \emph{Primordial black
  holes and their gravitational-wave signatures},  (2023)
  [\href{https://arxiv.org/abs/2310.19857}{{\ttfamily 2310.19857}}].

\bibitem{Pattison:2017mbe}
C.~Pattison, V.~Vennin, H.~Assadullahi and D.~Wands, \emph{Quantum diffusion
  during inflation and primordial black holes},
  \href{https://doi.org/10.1088/1475-7516/2017/10/046}{\emph{Journal of
  Cosmology and Astroparticle Physics} {\bfseries 2017}{\bfseries (10)} (2017)
  046} [\href{https://arxiv.org/abs/1707.00537}{{\ttfamily 1707.00537}}].

\bibitem{Starobinsky:1986fx}
A.~A. Starobinsky,
  \href{https://doi.org/10.1007/3-540-16452-9\_6}{\emph{{Stochastic De Sitter
  (Inflationary) Stage in the Early Universe}}, } in \emph{Lecture Notes in
  Physics (Field Theory, Quantum Gravity and Strings)}, vol.~246, p.~107, 1988.

\bibitem{Starobinsky:1994bd}
A.~A. Starobinsky and J.~Yokoyama, \emph{{Equilibrium state of a
  self-interacting scalar field in the de Sitter background}},
  \href{https://doi.org/10.1103/PhysRevD.50.6357}{\emph{Physical Review D}
  {\bfseries 50}{\bfseries (10)} (1994) 6357}
  [\href{https://arxiv.org/abs/astro-ph/9407016}{{\ttfamily
  astro-ph/9407016}}].

\bibitem{Starobinsky:1986fxa}
A.~A. Starobinskii, \emph{{Multicomponent de Sitter (inflationary) stages and
  the generation of perturbations}},
  \href{http://www.jetpletters.ru/ps/1419/article_21563.shtml}{\emph{Journal of
  Experimental and Theoretical Physics Letters} {\bfseries 42}{\bfseries (3)}
  (1985) 152}.

\bibitem{Sasaki1996}
M.~Sasaki and E.~D. Stewart, \emph{{A General Analytic Formula for the Spectral
  Index of the Density Perturbations Produced during Inflation}},
  \href{https://doi.org/10.1143/PTP.95.71}{\emph{Progress of Theoretical
  Physics} {\bfseries 95}{\bfseries (1)} (1996) 71}
  [\href{https://arxiv.org/abs/astro-ph/9507001}{{\ttfamily
  astro-ph/9507001}}].

\bibitem{Sasaki:1998ug}
M.~Sasaki and T.~Tanaka, \emph{{Super-Horizon Scale Dynamics of Multi-Scalar
  Inflation}}, \href{https://doi.org/10.1143/PTP.99.763}{\emph{Progress of
  Theoretical Physics} {\bfseries 99}{\bfseries (5)} (1998) 763}
  [\href{https://arxiv.org/abs/gr-qc/9801017}{{\ttfamily gr-qc/9801017}}].

\bibitem{Lyth:2004gb}
D.~H. Lyth, K.~A. Malik and M.~Sasaki, \emph{A general proof of the
  conservation of the curvature perturbation},
  \href{https://doi.org/10.1088/1475-7516/2005/05/004}{\emph{Journal of
  Cosmology and Astroparticle Physics} {\bfseries 2005}{\bfseries (2005)}
  (2005) 004} [\href{https://arxiv.org/abs/astro-ph/0411220}{{\ttfamily
  astro-ph/0411220}}].

\bibitem{Vennin:2015hra}
V.~Vennin and A.~A. Starobinsky, \emph{Correlation functions in stochastic
  inflation}, \href{https://doi.org/10.1140/epjc/s10052-015-3643-y}{\emph{The
  European Physical Journal C} {\bfseries 75} (2015) 413}
  [\href{https://arxiv.org/abs/1506.04732}{{\ttfamily 1506.04732}}].

\bibitem{Ezquiaga:2019ftu}
J.~M. Ezquiaga, J.~García-Bellido and V.~Vennin, \emph{The exponential tail of
  inflationary fluctuations: consequences for primordial black holes},
  \href{https://doi.org/10.1088/1475-7516/2020/03/029}{\emph{Journal of
  Cosmology and Astroparticle Physics} {\bfseries 2020}{\bfseries (03)} (2020)
  029} [\href{https://arxiv.org/abs/1912.05399}{{\ttfamily 1912.05399}}].

\bibitem{Animali:2022otk}
C.~Animali and V.~Vennin, \emph{Primordial black holes from stochastic
  tunnelling},
  \href{https://doi.org/10.1088/1475-7516/2023/02/043}{\emph{Journal of
  Cosmology and Astroparticle Physics} {\bfseries 2023}{\bfseries (02)} (2023)
  043} [\href{https://arxiv.org/abs/2210.03812}{{\ttfamily 2210.03812}}].

\bibitem{Briaud:2023eae}
V.~Briaud and V.~Vennin, \emph{Uphill inflation},
  \href{https://doi.org/10.1088/1475-7516/2023/06/029}{\emph{Journal of
  Cosmology and Astroparticle Physics} {\bfseries 2023}{\bfseries (06)} (2023)
  029} [\href{https://arxiv.org/abs/2301.09336}{{\ttfamily 2301.09336}}].

\bibitem{Figueroa:2020jkf}
D.~G. Figueroa, S.~Raatikainen, S.~Räsänen and E.~Tomberg,
  \emph{{Non-Gaussian Tail of the Curvature Perturbation in Stochastic
  Ultraslow-Roll Inflation: Implications for Primordial Black Hole
  Production}},
  \href{https://doi.org/10.1103/PhysRevLett.127.101302}{\emph{Physical Review
  Letters} {\bfseries 127}{\bfseries (10)} (2021) 101302}
  [\href{https://arxiv.org/abs/2012.06551}{{\ttfamily 2012.06551}}].

\bibitem{Figueroa:2021zah}
D.~G. Figueroa, S.~Raatikainen, S.~Räsänen and E.~Tomberg, \emph{Implications
  of stochastic effects for primordial black hole production in ultra-slow-roll
  inflation},
  \href{https://doi.org/10.1088/1475-7516/2022/05/027}{\emph{Journal of
  Cosmology and Astroparticle Physics} {\bfseries 2022}{\bfseries (2022)}
  (2022) 027} [\href{https://arxiv.org/abs/2111.07437}{{\ttfamily
  2111.07437}}].

\bibitem{Mazonka:1998ge}
O.~Mazonka, C.~Jarzynski and J.~Blocki, \emph{{Computing probabilities of very
  rare events for Langevin processes: A New method based on importance
  sampling}}, \href{https://doi.org/10.1016/S0375-9474(98)00478-3}{\emph{Nucl.
  Phys. A} {\bfseries 641} (1998) 335}
  [\href{https://arxiv.org/abs/nucl-th/9809075}{{\ttfamily nucl-th/9809075}}].

\bibitem{Jackson:2022unc}
J.~H.~P. Jackson \textit{et~al.}, \emph{Numerical simulations of stochastic
  inflation using importance sampling},
  \href{https://doi.org/10.1088/1475-7516/2022/10/067}{\emph{Journal of
  Cosmology and Astroparticle Physics} {\bfseries 2022}{\bfseries (10)} (2022)
  067} [\href{https://arxiv.org/abs/2206.11234}{{\ttfamily 2206.11234}}].

\bibitem{Biagetti:2018pjj}
M.~Biagetti, G.~Franciolini, A.~Kehagias and A.~Riotto, \emph{Primordial black
  holes from inflation and quantum diffusion},
  \href{https://doi.org/10.1088/1475-7516/2018/07/032}{\emph{Journal of
  Cosmology and Astroparticle Physics} {\bfseries 2018}{\bfseries (2018)}
  (2018) 032} [\href{https://arxiv.org/abs/1804.07124}{{\ttfamily
  1804.07124}}].

\bibitem{Firouzjahi:2020jrj}
H.~Firouzjahi, A.~Nassiri-Rad and M.~Noorbala, \emph{Stochastic nonattractor
  inflation}, \href{https://doi.org/10.1103/PhysRevD.102.123504}{\emph{Physical
  Review D} {\bfseries 102}{\bfseries (12)} (2020) 123504}
  [\href{https://arxiv.org/abs/2009.04680}{{\ttfamily 2009.04680}}].

\bibitem{Cai:2022erk}
Y.-F. Cai \textit{et~al.}, \emph{{Highly non-Gaussian tails and primordial
  black holes from single-field inflation}},
  \href{https://doi.org/10.1088/1475-7516/2022/12/034}{\emph{Journal of
  Cosmology and Astroparticle Physics} {\bfseries 2022}{\bfseries (2022)}
  (2022) 034} [\href{https://arxiv.org/abs/2207.11910}{{\ttfamily
  2207.11910}}].

\bibitem{Pi:2022ysn}
S.~Pi and M.~Sasaki, \emph{{Logarithmic Duality of the Curvature
  Perturbation}},
  \href{https://doi.org/10.1103/PhysRevLett.131.011002}{\emph{Physical Review
  Letters} {\bfseries 131}{\bfseries (1)} (2023) 011002}
  [\href{https://arxiv.org/abs/2211.13932}{{\ttfamily 2211.13932}}].

\bibitem{Hooshangi:2023kss}
S.~Hooshangi, M.~H. Namjoo and M.~Noorbala, \emph{Tail diversity from
  inflation},  (2023) [\href{https://arxiv.org/abs/2305.19257}{{\ttfamily
  2305.19257}}].

\bibitem{Wang:2024vfv}
X.~Wang, Y.-l. Zhang and M.~Sasaki, \emph{Enhanced curvature perturbation and
  primordial black hole formation in two-stage inflation with a break},
  \href{https://doi.org/10.1088/1475-7516/2024/07/076}{\emph{Journal of
  Cosmology and Astroparticle Physics} {\bfseries 2024}{\bfseries (07)} (2024)
  076} [\href{https://arxiv.org/abs/2404.02492}{{\ttfamily 2404.02492}}].

\bibitem{Tomberg:2023kli}
E.~Tomberg, \emph{Stochastic constant-roll inflation and primordial black
  holes}, \href{https://doi.org/10.1103/PhysRevD.108.043502}{\emph{Physical
  Review D} {\bfseries 108}{\bfseries (4)} (2023) 043502}
  [\href{https://arxiv.org/abs/2304.10903}{{\ttfamily 2304.10903}}].

\bibitem{Ballesteros:2024pwn}
G.~Ballesteros \textit{et~al.}, \emph{{Non-Gaussian tails without stochastic
  inflation}},  (2024) [\href{https://arxiv.org/abs/2406.02417}{{\ttfamily
  2406.02417}}].

\bibitem{Jackson:2023obv}
J.~H.~P. Jackson \textit{et~al.}, \emph{The separate-universe approach and
  sudden transitions during inflation},
  \href{https://doi.org/10.1088/1475-7516/2024/05/053}{\emph{Journal of
  Cosmology and Astroparticle Physics} {\bfseries 2024}{\bfseries (05)} (2024)
  053} [\href{https://arxiv.org/abs/2311.03281}{{\ttfamily 2311.03281}}].

\bibitem{Artigas:2024xhc}
D.~Artigas, S.~Pi and T.~Tanaka, \emph{{Extended $\delta N$ formalism}},
  (2024) [\href{https://arxiv.org/abs/2408.09964}{{\ttfamily 2408.09964}}].

\bibitem{Dimopoulos:2017ged}
K.~Dimopoulos, \emph{{Ultra slow-roll inflation demystified}},
  \href{https://doi.org/10.1016/j.physletb.2017.10.066}{\emph{Phys. Lett. B}
  {\bfseries 775} (2017) 262}
  [\href{https://arxiv.org/abs/1707.05644}{{\ttfamily 1707.05644}}].

\bibitem{Bassett:2005xm}
B.~A. Bassett, S.~Tsujikawa and D.~Wands, \emph{Inflation dynamics and
  reheating}, \href{https://doi.org/10.1103/RevModPhys.78.537}{\emph{Reviews of
  Modern Physics} {\bfseries 78}{\bfseries (2)} (2006) 537}
  [\href{https://arxiv.org/abs/astro-ph/0507632}{{\ttfamily
  astro-ph/0507632}}].

\bibitem{Wands:2000dp}
D.~Wands, K.~A. Malik, D.~H. Lyth and A.~R. Liddle, \emph{New approach to the
  evolution of cosmological perturbations on large scales},
  \href{https://doi.org/10.1103/PhysRevD.62.043527}{\emph{Physical Review D}
  {\bfseries 62}{\bfseries (4)} (2000) 043527}
  [\href{https://arxiv.org/abs/astro-ph/0003278}{{\ttfamily
  astro-ph/0003278}}].

\bibitem{Pattison:2019hef}
C.~Pattison, V.~Vennin, H.~Assadullahi and D.~Wands, \emph{Stochastic inflation
  beyond slow roll},
  \href{https://doi.org/10.1088/1475-7516/2019/07/031}{\emph{Journal of
  Cosmology and Astroparticle Physics} {\bfseries 2019}{\bfseries (2019)}
  (2019) 031} [\href{https://arxiv.org/abs/1905.06300}{{\ttfamily
  1905.06300}}].

\bibitem{Kodama:1997qw}
H.~Kodama and T.~Hamazaki, \emph{{Evolution of cosmological perturbations in
  the long wavelength limit}},
  \href{https://doi.org/10.1103/PhysRevD.57.7177}{\emph{Phys. Rev. D}
  {\bfseries 57} (1998) 7177}
  [\href{https://arxiv.org/abs/gr-qc/9712045}{{\ttfamily gr-qc/9712045}}].

\bibitem{Gordon:2000hv}
C.~Gordon, D.~Wands, B.~A. Bassett and R.~Maartens, \emph{Adiabatic and entropy
  perturbations from inflation},
  \href{https://doi.org/10.1103/PhysRevD.63.023506}{\emph{Physical Review D}
  {\bfseries 63}{\bfseries (2)} (2001) 023506}
  [\href{https://arxiv.org/abs/astro-ph/0009131}{{\ttfamily
  astro-ph/0009131}}].

\bibitem{Romano:2015vxz}
A.~E. Romano, S.~Mooij and M.~Sasaki, \emph{Adiabaticity and gravity theory
  independent conservation laws for cosmological perturbations},
  \href{https://doi.org/10.1016/j.physletb.2016.02.054}{\emph{Physics Letters
  B} {\bfseries 755} (2016) 464}
  [\href{https://arxiv.org/abs/1512.05757}{{\ttfamily 1512.05757}}].

\bibitem{Malik:2008im}
K.~A. Malik and D.~Wands, \emph{Cosmological perturbations},
  \href{https://doi.org/10.1016/j.physrep.2009.03.001}{\emph{Physics Reports}
  {\bfseries 475}{\bfseries (1--4)} (2009) 1}
  [\href{https://arxiv.org/abs/0809.4944}{{\ttfamily 0809.4944}}].

\bibitem{Inui:2024sce}
R.~Inui \textit{et~al.}, \emph{{Constant roll and non-Gaussian tail in light of
  logarithmic duality}},  (2024)
  [\href{https://arxiv.org/abs/2409.13500}{{\ttfamily 2409.13500}}].

\bibitem{Polarski1996}
D.~Polarski and A.~A. Starobinsky, \emph{Semiclassicality and decoherence of
  cosmological perturbations},
  \href{https://doi.org/10.1088/0264-9381/13/3/006}{\emph{Classical and Quantum
  Gravity} {\bfseries 13}{\bfseries (3)} (1996) 377}
  [\href{https://arxiv.org/abs/gr-qc/9504030}{{\ttfamily gr-qc/9504030}}].

\bibitem{Grain:2017dqa}
J.~Grain and V.~Vennin, \emph{{Stochastic inflation in phase space: Is slow
  roll a stochastic attractor?}},
  \href{https://doi.org/10.1088/1475-7516/2017/05/045}{\emph{Journal of
  Cosmology and Astroparticle Physics} {\bfseries 2017}{\bfseries (05)} (2017)
  045} [\href{https://arxiv.org/abs/1703.00447}{{\ttfamily 1703.00447}}].

\bibitem{Pattison:2021oen}
C.~Pattison, V.~Vennin, D.~Wands and H.~Assadullahi, \emph{Ultra-slow-roll
  inflation with quantum diffusion},
  \href{https://doi.org/10.1088/1475-7516/2021/04/080}{\emph{Journal of
  Cosmology and Astroparticle Physics} {\bfseries 2021}{\bfseries (04)} (2021)
  080} [\href{https://arxiv.org/abs/2101.05741}{{\ttfamily 2101.05741}}].

\bibitem{Firouzjahi:2018vet}
H.~Firouzjahi, A.~Nassiri-Rad and M.~Noorbala, \emph{{Stochastic Ultra Slow
  Roll Inflation}},
  \href{https://doi.org/10.1088/1475-7516/2019/01/040}{\emph{Journal of
  Cosmology and Astroparticle Physics} {\bfseries 2019}{\bfseries (01)} (2019)
  040} [\href{https://arxiv.org/abs/1811.02175}{{\ttfamily 1811.02175}}].

\bibitem{Stewart:1993bc}
E.~D. Stewart and D.~H. Lyth, \emph{{A More accurate analytic calculation of
  the spectrum of cosmological perturbations produced during inflation}},
  \href{https://doi.org/10.1016/0370-2693(93)90379-V}{\emph{Physics Letters B}
  {\bfseries 302} (1993) 171}
  [\href{https://arxiv.org/abs/gr-qc/9302019}{{\ttfamily gr-qc/9302019}}].

\bibitem{De:2020hdo}
A.~De and R.~Mahbub, \emph{Numerically modeling stochastic inflation in
  slow-roll and beyond},
  \href{https://doi.org/10.1103/PhysRevD.102.123509}{\emph{Physical Review D}
  {\bfseries 102}{\bfseries (12)} (2020) 123509}
  [\href{https://arxiv.org/abs/2010.12685}{{\ttfamily 2010.12685}}].

\bibitem{Ahmadi:2022lsm}
N.~Ahmadi \textit{et~al.}, \emph{Quantum diffusion in sharp transition to
  non-slow-roll phase},
  \href{https://doi.org/10.1088/1475-7516/2022/08/078}{\emph{Journal of
  Cosmology and Astroparticle Physics} {\bfseries 2022}{\bfseries (2022)}
  (2022) 078} [\href{https://arxiv.org/abs/2207.10578}{{\ttfamily
  2207.10578}}].

\bibitem{Mishra:2023lhe}
S.~S. Mishra, E.~J. Copeland and A.~M. Green, \emph{{Primordial black holes and
  stochastic inflation beyond slow roll. Part I. Noise matrix elements}},
  \href{https://doi.org/10.1088/1475-7516/2023/09/005}{\emph{Journal of
  Cosmology and Astroparticle Physics} {\bfseries 2023}{\bfseries (09)} (2023)
  005} [\href{https://arxiv.org/abs/2303.17375}{{\ttfamily 2303.17375}}].

\bibitem{Kloeden1992}
P.~E. Kloeden and E.~Platen,
  \href{https://doi.org/10.1007/978-3-662-12616-5}{\emph{{Numerical Solution of
  Stochastic Differential Equations}}}, Springer Berlin Heidelberg (1992).

\bibitem{Starobinsky:1992ts}
A.~A. Starobinskii, \emph{Spectrum of adiabatic perturbations in the universe
  when there are singularities in the inflation potential},
  \href{http://www.jetpletters.ru/ps/1276/article_19291.shtml}{\emph{Journal of
  Experimental and Theoretical Physics Letters} {\bfseries 55}{\bfseries (9)}
  (1992) 489}.

\bibitem{Leach:2001zf}
S.~M. Leach, M.~Sasaki, D.~Wands and A.~R. Liddle, \emph{Enhancement of
  superhorizon scale inflationary curvature perturbations},
  \href{https://doi.org/10.1103/PhysRevD.64.023512}{\emph{Physical Review D}
  {\bfseries 64}{\bfseries (2)} (2001) 023512}
  [\href{https://arxiv.org/abs/astro-ph/0101406}{{\ttfamily
  astro-ph/0101406}}].

\bibitem{Martin:2011sn}
J.~Martin and L.~Sriramkumar, \emph{{The scalar bi-spectrum in the Starobinsky
  model: The equilateral case}},
  \href{https://doi.org/10.1088/1475-7516/2012/01/008}{\emph{Journal of
  Cosmology and Astroparticle Physics} {\bfseries 2012}{\bfseries (2012)}
  (2012) 008} [\href{https://arxiv.org/abs/1109.5838}{{\ttfamily 1109.5838}}].

\bibitem{Martin:2014kja}
J.~Martin, L.~Sriramkumar and D.~K. Hazra, \emph{Sharp inflaton potentials and
  bi-spectra: effects of smoothening the discontinuity},
  \href{https://doi.org/10.1088/1475-7516/2014/09/039}{\emph{Journal of
  Cosmology and Astroparticle Physics} {\bfseries 2014}{\bfseries (2014)}
  (2014) 039} [\href{https://arxiv.org/abs/1404.6093}{{\ttfamily 1404.6093}}].

\bibitem{Pi:2022zxs}
S.~Pi and J.~Wang, \emph{{Primordial black hole formation in Starobinsky's
  linear potential model}},
  \href{https://doi.org/10.1088/1475-7516/2023/06/018}{\emph{Journal of
  Cosmology and Astroparticle Physics} {\bfseries 2023}{\bfseries (2023)}
  (2023) 018} [\href{https://arxiv.org/abs/2209.14183}{{\ttfamily
  2209.14183}}].

\bibitem{Tomberg:2022mkt}
E.~Tomberg, \emph{Numerical stochastic inflation constrained by frozen noise},
  \href{https://doi.org/10.1088/1475-7516/2023/04/042}{\emph{Journal of
  Cosmology and Astroparticle Physics} {\bfseries 2023}{\bfseries (04)} (2023)
  042} [\href{https://arxiv.org/abs/2210.17441}{{\ttfamily 2210.17441}}].

\bibitem{Mishra:2019pzq}
S.~S. Mishra and V.~Sahni, \emph{Primordial black holes from a tiny bump/dip in
  the inflaton potential},
  \href{https://doi.org/10.1088/1475-7516/2020/04/007}{\emph{Journal of
  Cosmology and Astroparticle Physics} {\bfseries 2020}{\bfseries (2020)}
  (2020) 007} [\href{https://arxiv.org/abs/1911.00057}{{\ttfamily
  1911.00057}}].

\bibitem{Cole:2023wyx}
P.~S. Cole, A.~D. Gow, C.~T. Byrnes and S.~P. Patil, \emph{Primordial black
  holes from single-field inflation: a fine-tuning audit},
  \href{https://doi.org/10.1088/1475-7516/2023/08/031}{\emph{Journal of
  Cosmology and Astroparticle Physics} {\bfseries 2023}{\bfseries (2023)}
  (2023) 031} [\href{https://arxiv.org/abs/2304.01997}{{\ttfamily
  2304.01997}}].

\bibitem{Tada:2016pmk}
Y.~Tada and V.~Vennin, \emph{{Squeezed bispectrum in the $\delta N$ formalism:
  local observer effect in field space}},
  \href{https://doi.org/10.1088/1475-7516/2017/02/021}{\emph{Journal of
  Cosmology and Astroparticle Physics} {\bfseries 2017}{\bfseries (02)} (2017)
  021} [\href{https://arxiv.org/abs/1609.08876}{{\ttfamily 1609.08876}}].

\bibitem{Tada:2021zzj}
Y.~Tada and V.~Vennin, \emph{Statistics of coarse-grained cosmological fields
  in stochastic inflation},
  \href{https://doi.org/10.1088/1475-7516/2022/02/021}{\emph{Journal of
  Cosmology and Astroparticle Physics} {\bfseries 2022}{\bfseries (02)} (2022)
  021} [\href{https://arxiv.org/abs/2111.15280}{{\ttfamily 2111.15280}}].

\bibitem{Ando:2020fjm}
K.~Ando and V.~Vennin, \emph{Power spectrum in stochastic inflation},
  \href{https://doi.org/10.1088/1475-7516/2021/04/057}{\emph{Journal of
  Cosmology and Astroparticle Physics} {\bfseries 2021}{\bfseries (04)} (2021)
  057} [\href{https://arxiv.org/abs/2012.02031}{{\ttfamily 2012.02031}}].

\bibitem{Animali:2024jiz}
C.~Animali and V.~Vennin, \emph{Clustering of primordial black holes from
  quantum diffusion during inflation},
  \href{https://doi.org/10.1088/1475-7516/2024/08/026}{\emph{Journal of
  Cosmology and Astroparticle Physics} {\bfseries 2024}{\bfseries (08)} (2024)
  026} [\href{https://arxiv.org/abs/2402.08642}{{\ttfamily 2402.08642}}].

\end{thebibliography}\endgroup

\end{document}